\documentclass[ALICE,manyauthors]{cernphprep}
\usepackage[comma,square,numbers,sort&compress]{natbib}
\usepackage{hyperref}
\usepackage{booktabs}
\usepackage{lineno}
\usepackage{color}
\usepackage{cleveref} %Multiple references
\usepackage[T1]{fontenc}
\usepackage{orcidlink}
\newcolumntype{Y}{>{\centering\arraybackslash}X}
\RequirePackage[normalem]{ulem} %DIF PREAMBLE
\RequirePackage{color}\definecolor{RED}{rgb}{1,0,0}\definecolor{BLUE}{rgb}{0,0,1} %DIF PREAMBLE
\begin{document}
%%%%%%%%%%%%%%%%%%%%%%%%%%%%%%%%%%%%%%%%%%%%%%%%%%
% These are some new commands that may be useful 
% for paper writing in general. If other newcommands
% are needed for your specific paper, please feel 
% free to add here. 
%
% The currently available commands are organized in: 
% 1) Systems
% 2) Quantities
% 3) Energies and units
% 4) Detectors
% 5) particle species 
%%%%%%%%%%%%%%%%%%%%%%%%%%%%%%%%%%%%%%%%%%%%%%%%%%

%%%%%%%%%%%%%%%  Command definition %%%%%%%%%%%%%%%%%%%%%%%%
\newcommand{\RT}{\ensuremath{R_{\rm T}}\xspace}
\newcommand{\Nm}{\ensuremath{N_{\mathrm{m}}}\xspace}
\newcommand{\Nt}{\ensuremath{N_{\mathrm{t}}}\xspace}
\newcommand{\NT}{\ensuremath{N_{\mathrm{T}}}\xspace}
\newcommand{\mNT}{\ensuremath{\langle N_{\mathrm{T}}} \rangle \xspace}
\newcommand{\Smt}{\ensuremath{S_{\mathrm{mt}}}\xspace}

\newcommand{\pipm}{\ensuremath{\pi^{\pm}}\xspace}
\newcommand{\kpm}{\ensuremath{\mathrm{K}^{\pm}}\xspace}
\newcommand{\ppm}{\ensuremath{(\overline{\mathrm{p}})\mathrm{p}}\xspace}

% Unfolding

\newcommand{\RMp}{\ensuremath{P(N_{\mathrm{T,m}}|N_{\mathrm{T,t}})}\xspace}
\newcommand{\NTc}{\ensuremath{N_{\mathrm{T}}}\xspace}
\newcommand{\NTt}{\ensuremath{N_{\mathrm{T,t}}}\xspace}
\newcommand{\NTm}{\ensuremath{N_{\mathrm{T,m}}}\xspace}
\newcommand{\UM}{\ensuremath{M_{\mathrm{tm}}}\xspace}

\newcommand{\UnfDis}{\ensuremath{Y(N_{\mathrm{T,t}}})\xspace}
\newcommand{\RawDis}{\ensuremath{Y(N_{\mathrm{T,m}}})\xspace}
\newcommand{\Mone}{\ensuremath{\mathrm{M1}_{\mathrm{tm}}}\xspace}
\newcommand{\Mtwo}{\ensuremath{\mathrm{M2}_{\mathrm{tm}}(p_{\mathrm{T}})}\xspace}
\newcommand{\Prior}{\ensuremath{P_{0}(\NTm)}\xspace}
\newcommand{\UpdatedPrior}{\ensuremath{\widehat{P}(\NTt)}\xspace}
\newcommand{\RawYield}{\ensuremath{\mathrm{d}Y(N_{\mathrm{T,m}})/\mathrm{d}p_{\mathrm{T}}}}
\newcommand{\UnfYield}{\ensuremath{\mathrm{d}Y(N_{\mathrm{T,t}})/\mathrm{d}p_{\mathrm{T}}}}

% 1) SYSTEMS 
\newcommand{\pp}           {pp\xspace}
\newcommand{\ppbar}        {\mbox{$\mathrm {p\overline{p}}$}\xspace}
\newcommand{\XeXe}         {\mbox{Xe--Xe}\xspace}
\newcommand{\PbPb}         {\mbox{Pb--Pb}\xspace}
\newcommand{\pA}           {\mbox{pA}\xspace}
\newcommand{\pPb}          {\mbox{p--Pb}\xspace}
\newcommand{\AuAu}         {\mbox{Au--Au}\xspace}
\newcommand{\dAu}          {\mbox{d--Au}\xspace}

% 2) QUANTITIES 
\newcommand{\s}            {\ensuremath{\sqrt{s}}\xspace}
\newcommand{\snn}          {\ensuremath{\sqrt{s_{\mathrm{NN}}}}\xspace}
\newcommand{\pt}           {\ensuremath{p_{\rm T}}\xspace}
\newcommand{\ptleading}    {\ensuremath{p_{\rm T}^{\rm{leading}}}\xspace}
\newcommand{\meanpt}       {\ensuremath{\langle p_{\rm T}\rangle}\xspace}
\newcommand{\ycms}         {\ensuremath{y_{\rm CMS}}\xspace}
\newcommand{\ylab}         {\ensuremath{y_{\rm lab}}\xspace}
\newcommand{\etarange}[1]  {\mbox{$\left | \eta \right |~<~#1$}}
\newcommand{\yrange}[1]    {\mbox{$\left | y \right |~<~#1$}}
\newcommand{\dndy}         {\ensuremath{\mathrm{d}N_\mathrm{ch}/\mathrm{d}y}\xspace}
\newcommand{\dndeta}       {\ensuremath{\mathrm{d}N_\mathrm{ch}/\mathrm{d}\eta}\xspace}
\newcommand{\avdndeta}     {\ensuremath{\langle\dndeta\rangle}\xspace}
\newcommand{\dNdy}         {\ensuremath{\mathrm{d}N_\mathrm{ch}/\mathrm{d}y}\xspace}
\newcommand{\Npart}        {\ensuremath{N_\mathrm{part}}\xspace}
\newcommand{\Ncoll}        {\ensuremath{N_\mathrm{coll}}\xspace}
\newcommand{\dEdx}         {\ensuremath{\textrm{d}E/\textrm{d}x}\xspace}
\newcommand{\meandEdx}{\ensuremath{\langle\textrm{d}E/\textrm{d}x}\rangle\xspace}

\newcommand{\RpPb}         {\ensuremath{R_{\rm pPb}}\xspace}

% 3) ENERGIES, UNITS
\newcommand{\nineH}        {$\sqrt{s}~=~0.9$~Te\kern-.1emV\xspace}
\newcommand{\seven}        {$\sqrt{s}~=~7$~Te\kern-.1emV\xspace}
\newcommand{\twoH}         {$\sqrt{s}~=~0.2$~Te\kern-.1emV\xspace}
\newcommand{\twosevensix}  {$\sqrt{s}~=~2.76$~Te\kern-.1emV\xspace}
\newcommand{\five}         {$\sqrt{s}~=~5.02$~Te\kern-.1emV\xspace}
\newcommand{\thirteen}    {$\sqrt{s}~=~13$~Te\kern-.1emV\xspace}
\newcommand{\twosevensixnn}{$\sqrt{s_{\mathrm{NN}}}~=~2.76$~Te\kern-.1emV\xspace}
\newcommand{\fivenn}       {$\sqrt{s_{\mathrm{NN}}}~=~5.02$~Te\kern-.1emV\xspace}
\newcommand{\LT}           {L{\'e}vy-Tsallis\xspace}
\newcommand{\GeVc}         {Ge\kern-.1emV/$c$\xspace}
\newcommand{\MeVc}         {Me\kern-.1emV/$c$\xspace}
\newcommand{\TeV}          {Te\kern-.1emV\xspace}
\newcommand{\GeV}          {Ge\kern-.1emV\xspace}
\newcommand{\MeV}          {Me\kern-.1emV\xspace}
\newcommand{\GeVmass}      {Ge\kern-.2emV/$c^2$\xspace}
\newcommand{\MeVmass}      {Me\kern-.2emV/$c^2$\xspace}
\newcommand{\lumi}         {\ensuremath{\mathcal{L}}\xspace}
\newcommand{\gevc}         {\ensuremath{\mathrm{GeV}/c}\xspace}

% 4) DETECTORS 
\newcommand{\ITS}          {\rm{ITS}\xspace}
\newcommand{\TOF}          {\rm{TOF}\xspace}
\newcommand{\ZDC}          {\rm{ZDC}\xspace}
\newcommand{\ZDCs}         {\rm{ZDCs}\xspace}
\newcommand{\ZNA}          {\rm{ZNA}\xspace}
\newcommand{\ZNC}          {\rm{ZNC}\xspace}
\newcommand{\SPD}          {\rm{SPD}\xspace}
\newcommand{\SDD}          {\rm{SDD}\xspace}
\newcommand{\SSD}          {\rm{SSD}\xspace}
\newcommand{\TPC}          {\rm{TPC}\xspace}
\newcommand{\TRD}          {\rm{TRD}\xspace}
\newcommand{\VZERO}        {\rm{V0}\xspace}
\newcommand{\VZEROA}       {\rm{V0A}\xspace}
\newcommand{\VZEROC}       {\rm{V0C}\xspace}
\newcommand{\Vdecay} 	   {\ensuremath{V^{0}}\xspace}

% 4) PARTICLE SPECIES 
\newcommand{\pion}         {\ensuremath{\pi}\xspace}
\newcommand{\kaon}         {\ensuremath{\textrm{K}}\xspace}
\newcommand{\pr}           {\ensuremath{\textrm{p}}\xspace}
\newcommand{\ee}           {\ensuremath{e^{+}e^{-}}} 
\newcommand{\pip}          {\ensuremath{\pi^{+}}\xspace}
\newcommand{\pim}          {\ensuremath{\pi^{-}}\xspace}
\newcommand{\kapm}         {\ensuremath{\mathrm{K}^{\pm}}}
\newcommand{\kap}          {\ensuremath{\rm{K}^{+}}\xspace}
\newcommand{\kam}          {\ensuremath{\rm{K}^{-}}\xspace}
\newcommand{\pbar}         {\ensuremath{\rm\overline{p}}\xspace}
\newcommand{\kzero}        {\ensuremath{{\rm K}^{0}_{\rm{S}}}\xspace}
\newcommand{\lmb}          {\ensuremath{\Lambda}\xspace}
\newcommand{\almb}         {\ensuremath{\overline{\Lambda}}\xspace}
\newcommand{\Om}           {\ensuremath{\Omega^-}\xspace}
\newcommand{\Mo}           {\ensuremath{\overline{\Omega}^+}\xspace}
\newcommand{\X}            {\ensuremath{\Xi^-}\xspace}
\newcommand{\Ix}           {\ensuremath{\overline{\Xi}^+}\xspace}
\newcommand{\Xis}          {\ensuremath{\Xi^{\pm}}\xspace}
\newcommand{\Oms}          {\ensuremath{\Omega^{\pm}}\xspace}
\newcommand{\degree}       {\ensuremath{^{\rm o}}\xspace}

% Particle ratios
\newcommand{\ktopi}        {\ensuremath{\mathrm{K}/\pi}\xspace}
\newcommand{\ptopi}        {\ensuremath{\mathrm{p}/\pi}\xspace}

% PID 
\newcommand{\nsigma}       {\ensuremath{\mathrm{n}_{\sigma}}\xspace}

%%%%%%%%%%%%%%%  Title page %%%%%%%%%%%%%%%%%%%%%%%%
\begin{titlepage}
% the dates below correspond to CERN approval
% please don't touch: EB chairs will take care
\PHyear{2022}       % required, will be obtained from CERN
\PHnumber{286}      % required, will be obtained from CERN
\PHdate{16 December}  % required, will be obtained from CERN
%%%%%%%%%%%%%%%%%%%%%%%%%%%%%%%%%%%%%%%%%%%%%%%%%%%%

%%% Put your own title + short title here:
\title{Production of pions, kaons, and protons as a
function of the relative transverse activity classifier in pp collisions at $\mathbf{\sqrt{{\textit s}}}=13$~\textbf{TeV}}
\ShortTitle{Production of \pion, \kaon and \pr as a function of \RT in pp collisions at $\sqrt{s}=13$ TeV}   % appears on left page headers

%%% Do not change the next lines
\Collaboration{ALICE Collaboration\thanks{See Appendix~\ref{app:collab} for the list of collaboration members}}
\ShortAuthor{ALICE Collaboration} % appears on right page headers, do not change

\begin{abstract}
The production of $\pi^{\pm}$, $\mathrm{K}^{\pm}$, and $(\pbar)\mathrm{p}$ is measured in \pp collisions at $\sqrt{s}\,= 13~\mathrm{TeV}$ in different topological regions of the events. Particle transverse momentum $(\pt)$ spectra are measured in the ``toward'', ``transverse'', and ``away'' angular regions defined with respect to the direction of the leading particle in the event. While the toward and away regions contain the fragmentation products of the near-side and away-side jets, respectively, the transverse region is dominated by particles from the Underlying Event (UE). The relative transverse activity classifier, $\RT = \NT / \langle \NT \rangle$, is used to group events according to their UE activity, where \NT is the measured charged-particle multiplicity per event in the transverse region and $\langle \NT \rangle$ is the mean value over all the analysed events. The first measurements of identified particle \pt spectra as a function of \RT in the three topological regions are reported. It is found that the yield of high transverse momentum particles relative to the \RT-integrated measurement decreases with increasing \RT in both the toward and the away regions, indicating that the softer UE dominates particle production as \RT increases and validating that \RT can be used to control the magnitude of the UE. Conversely, the spectral shapes in the transverse region harden significantly with increasing \RT. This hardening follows a mass ordering, being more significant for heavier particles. Finally, it is observed that the \pt-differential particle ratios $(\mathrm{p} + \pbar)/(\pip + \pim)$ and $(\mathrm{K}^{+} + \mathrm{K}^{-})/(\pip + \pim)$ in the low UE limit $(\RT \rightarrow 0)$ approach expectations from Monte Carlo generators such as PYTHIA 8 with Monash 2013 tune and EPOS LHC, where the jet-fragmentation models have been tuned to reproduce $\mathrm{e}^{+}\mathrm{e}^{-}$ results.
\end{abstract}

\end{titlepage}

\setcounter{page}{2} %please do not remove this line

%%%%%%%%%%%%%%%%%%%%%%%%%%%%%%%%
% begin main text
%%%%%%%%%%%%%%%%%%%%%%%%%%%%%%%%

\section{Introduction} 
In recent years, proton--proton (\pp) and proton--lead (\pPb) collisions, commonly denoted as small collision systems, have attracted the heavy-ion community's attention due to several measurements in high-multiplicity pp and p--Pb collisions, which show similar features as those observed in heavy-ion collisions. Observations of radial~\cite{CMS:2016zzh, Mult_dependence_hadrons_7TeV_pp,pikp_vs_mult_13TeV,Mult_depdencen_piKpL_5TeV_pPb} and anisotropic~\cite{CMS:2016fnw,ALICE:2019zfl} flows (collective phenomena), as well as strangeness enhancement~\cite{ALICE:2019avo,CMS:2016zzh,ALICE:2016fzo} in heavy-ion collisions, are associated with the formation of the strongly interacting quark--gluon plasma (QGP). However, these signatures have also been observed in \pp and \pPb collisions~\citep{Mult_depdencen_piKpL_5TeV_pPb,Mult_dependence_hadrons_7TeV_pp,ALICE:2016fzo,CMS:2017eoq}. In particular, the \pt-differential baryon-to-meson ratios in small collision systems showcase radial-flow like effects when studied as a function of the charged particle multiplicity of the event~\cite{Mult_dependence_hadrons_7TeV_pp,pikp_vs_mult_13TeV}. In order to pin down the origins of the effects observed in small collision systems, it has been proposed to study particle production as a function of the Underlying Event (UE) activity~\cite{Martin:2016igp}. The UE is defined as the particles that do not originate from the fragmentation products of the partons produced in the hardest scattering. It consists of the set of particles arising from initial- and final-state radiation, beam remnants and multiple parton interactions (MPIs)~\cite{PhysRevD.36.2019}. In the context of MPI models, the measurement of identified particle yields and ratios as a function of the UE activity allows one to measure event properties in an MPI-suppressed\,(-enhanced) environment. Moreover, as shown in~\cite{Ortiz:2016kpz}, these measurements can also provide insights into possible effects that give similar signatures as radial flow but are produced by jet hardening with increasing multiplicity.

At the LHC energies, particles and anti-particles are produced in equal amounts~\cite{ALICE:2013yba}. In the remaining of this paper and unless stated otherwise, the notation \pion, \kaon and \pr is adopted to refer to $(\pip + \pim)$, $(\mathrm{K}^{+} + \mathrm{K}^{-})$, and $(\mathrm{p}+\pbar)$, respectively. In this study, the production of \pion, \kaon, and \pr is studied as a function of the UE activity in \pp collisions at centre-of-mass energy, $\sqrt{s}=13~\mathrm{TeV}$. The UE is examined using the event topology defined by the leading charged particle in the event, which is defined as the charged particle with the highest transverse momentum in the range $5 \leq \ptleading < 40~\gevc$, and reconstructed in the pseudorapidity interval $|\eta|<0.8$. The lower \ptleading threshold corresponds to the onset of the UE plateau in the transverse region (transverse to the direction of the leading particle)~\cite{ALICE:2019mmy,CMS:2015yqj,ATLAS:2014yqy,CMS:2013ycn}. In the plateau region, quantities such as the average charged-particle density, $\langle N_{\textrm{ch}} \rangle$, and the average transverse momentum sum, $\langle \sum \pt \rangle$, have little dependence on the \pt of the leading particle or jet. This study uses a lower threshold on the \ptleading of $5~\mathrm{GeV}/c$ to guarantee that the multiple soft scatterings that contribute to the UE are largely independent of the \ptleading. In~\cite{ATLAS_UE_InJetEvents} a slow rise of the UE plateau is reported. This can be explained by additional contributions from wide-angle radiation associated with the hard scattering. Since wide-angle contamination becomes significant for jet $\pt > 50 ~\gevc$~\cite{ATLAS_UE_InJetEvents}, an upper limit on \ptleading of $40~\gevc$ is used to reduce its effects.  

\begin{figure}[!ht]
    \centering
    \hspace{0cm}
    \includegraphics[width=0.5\textwidth]{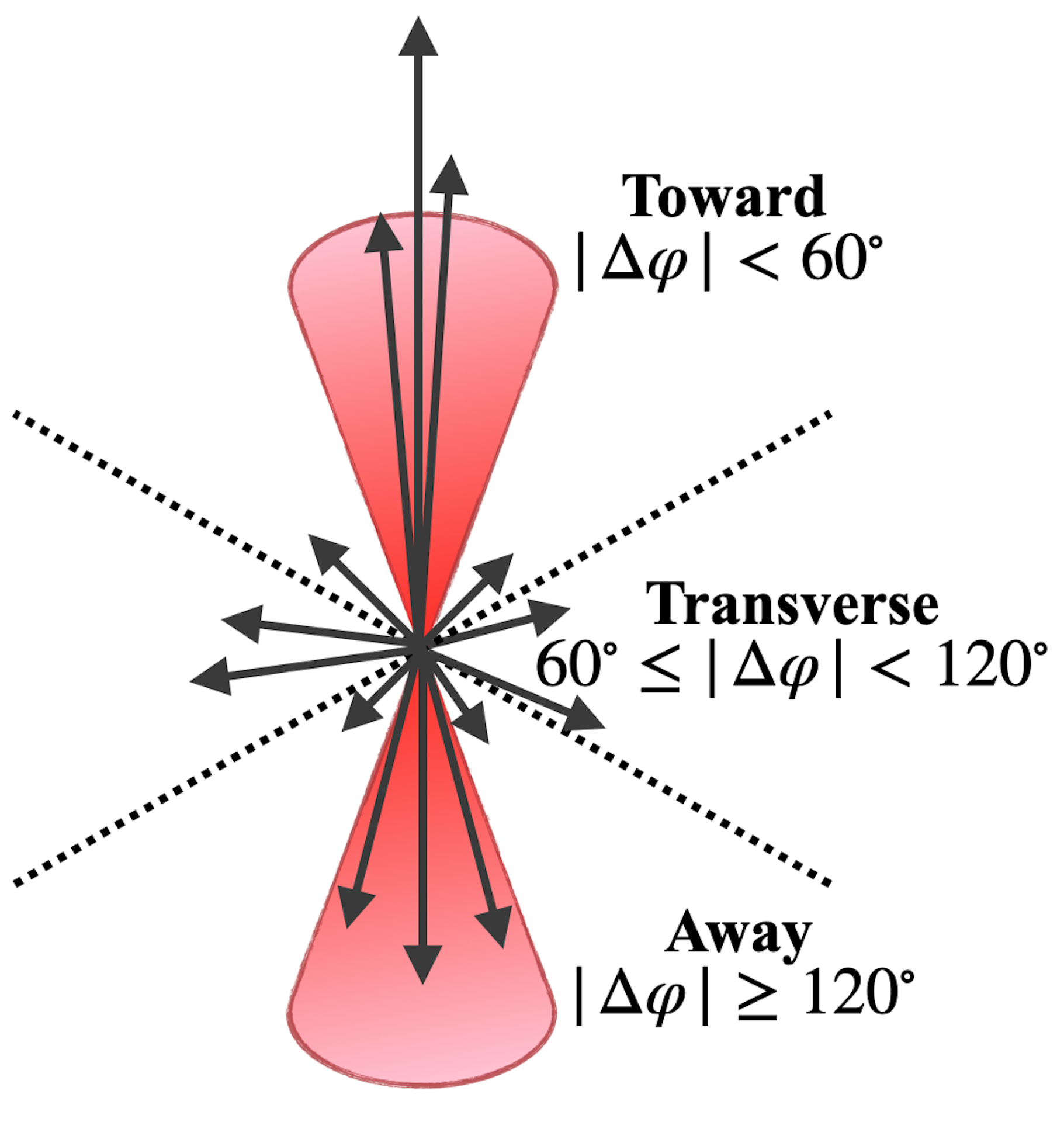}
    \caption{Illustration of the toward, transverse, and away regions in the azimuthal angle plane with respect to the direction of the leading particle. The leading particle is represented with the longest upright arrow. The UE is represented with the small arrows transverse to the leading particle. The red cones represent the jet and away-side jet.}
    \label{fig:azimuth_regions}
\end{figure}

To study the particle production associated with different underlying physics mechanisms, the conventional division of the azimuthal $(\varphi)$ plane into regions relative to the direction of the leading particle~\cite{Charged_Jet_evolution_and_UE_pbarp_1.8TeV} is used (see Fig.~\ref{fig:azimuth_regions}). The observables reported in this paper are measured in three different topological regions, the toward, transverse, and away regions. These are defined based on the absolute difference in azimuthal angle between the leading and associated particles, $|\Delta \varphi| = |\varphi^{\mathrm{leading}} - \varphi|$. The associated particles are measured in the kinematic range $0.15 \leq \pt < 5~\gevc$ and \etarange{0.8}. The toward, transverse, and away regions are defined by $| \Delta \varphi | < 60^{\circ}$, $ 60^{\circ} \leq | \Delta \varphi | < 120^{\circ}$, and $ | \Delta\varphi | \geq 120^{\circ}$, respectively. The particle production in the toward and away regions contains the constituents of the leading and away-side jets, respectively, the transverse region is mainly sensitive to multiple parton interactions and initial- and final-state radiations. 

The UE activity is quantified using the relative transverse activity classifier $\RT$~\cite{Martin:2016igp}, which is defined as $\NT / \langle \NT \rangle$, where \NT is the measured charged-particle multiplicity per event in the transverse region and $\langle \NT \rangle$ is the mean value over all the analysed events. By construction, \RT cleanly separates events with ``higher-than-average'' UE from ``lower-than-average'' ones irrespective of the centre-of-mass energy. Of particular interest is whether events with very low UE activity, which are dominated by the jet activity, exhibit particle ratios and spectra consistent with fragmentation models tuned to $\mathrm{e}^{+}\mathrm{e}^{-}$ data and whether events with high UE activity exhibit any clear signs of flow or other collective effects~\cite{Martin:2016igp}. Finally, it is worth mentioning that this study is complementary to the measurements made using transverse spherocity, in which global event properties are studied for jet-like and isotropic topologies~\cite{Ortiz:2015ttf}\cite{ALICE:2019dfi}.

The structure of the paper is as follows: In Sec.~\ref{sec:data_analysis}, the data analysis is described, Sec.~\ref{sec:systematic_uncertainties} discusses the systematic uncertainties, and in Sec.~\ref{sec:results}, the results are presented. Finally, in Sec.~\ref{sec:Conclusions}, the conclusions are given.

\section{Analysis procedure}\label{sec:data_analysis}

\subsection{Event and track selection}
\label{subsec:event_selection}

This study was carried out with the data collected in pp collisions at \thirteen by the ALICE Collaboration during the LHC runs from 2016 and 2018. A detailed description of the ALICE apparatus and its performance can be found in~\cite{ALICE:2008ngc,ALICE:2014sbx}. The subdetectors used in this analysis are the V0~\cite{Abbas:2013taa}, the Inner Tracking System (\ITS)~\cite{ALICE:2010tia}, the Time Projection Chamber (\TPC)~\cite{Alme:2010ke}, and the Time-Of-Flight (\TOF)~\cite{Akindinov:2013tea}. These subdetectors are located inside a $B=0.5~\mathrm{T}$ solenoidal magnetic field. The V0 detector consists of two arrays of 32 scintillators each, covering the forward $(\mathrm{V0A},~2.8 < \eta < 5.1)$ and backward $(\mathrm{V0C},~ -3.7 < \eta < -1.7)$ pseudorapidity regions. The \ITS is the innermost barrel detector. It consists of six cylindrical layers of high-resolution silicon tracking detectors: the two innermost layers of the Silicon Pixel Detector (\SPD) provide a digital readout and are also used as a trigger detector. The Silicon Drift Detector (\SDD) and the Silicon Strip Detector (\SSD) compose the four outer layers of the \ITS. Together, they provide the amplitude of the charge signal, which is used for particle identification through the measurement of the specific energy loss $(\dEdx)$. The \TPC is the primary detector for tracking and particle identification. It is a large cylindrical drift detector with a diameter and length of about $5\,\mathrm{m}$, which covers the pseudorapidity range $|\eta|<0.8$ with full-azimuth coverage. Particle identification is accomplished via the measurement of the \dEdx. In pp collisions the resolution of the \dEdx is about 5\,\%. The \TOF is a large-area array of multigap resistive plate chambers (MRPC), which surrounds the interaction point and covers the pseudorapidity region $|\eta| < 0.9$ with full-azimuth coverage. The time-of-flight is measured as the difference between the particle arrival time and the event collision time.

The event selection in this study follows those  of the previous studies to measure the production of \pion, \kaon, and \pr as a function of the charged-particle multiplicity in~\cite{pikp_vs_mult_PbPb_5TeV, pikp_vs_mult_13TeV}. The minimum-bias trigger requires signals in both \VZEROA and \VZEROC scintillators in coincidence with the arrival of the proton bunches from both directions. The primary vertex position is reconstructed using global tracks (reconstructed using \ITS and \TPC information). For events with too few tracks to compute the vertex position, the primary vertex from \SPD tracklets (reconstructed using only \SPD information) is used instead. Events are required to have a vertex position along the $z$-axis (parallel to the beam axis)  in $|z| < 10~\rm{cm}$, where $z = 0$ corresponds to the centre of the detector. The out-of-bunch pileup is rejected offline using the timing information from the two V0 subdetectors. Furthermore, events with multiple interaction vertices reconstructed are rejected. Finally, events are required to have a leading particle with $5 \leq \ptleading < 40~\mathrm{GeV}/c$. The total number of events after event and vertex selections amounts to about $827$ million, while the number of analysed events with a leading particle is about $8.1$ million.

The distributions presented in this study correspond to primary charged particles, which are defined as particles with a mean proper lifetime $\tau$ larger than $1 ~ \mathrm{cm}/c$, which are either produced directly in the interaction or from decays of particles with $\tau$ smaller than $1 ~ \mathrm{cm}/c$, excluding particles produced in interactions with material ~\cite{ALICE_PUBLIC_2017_005}. Primary charged particles are reconstructed using the \ITS and \TPC detectors, which provide measurements of the track transverse momentum and azimuthal angle. In particular, tracks are required to cross at least $70$ TPC pad rows. They are also required to have at least two hits in the ITS, out of which at least one is in the \SPD layers. The fit quality for the \ITS and \TPC track points must satisfy $\chi^{2}_{\ITS}/N_{\mathrm{hits}} < 36$ and $\chi^{2}_{\TPC}/N_{\mathrm{clusters}} < 4$, respectively, where $N_{\mathrm{hits}}$ and $N_{\mathrm{clusters}}$ are the number of hits in the \ITS and the number of clusters in the \TPC associated to the track, respectively. Finally, tracks are also required to have a transverse momentum larger than $0.15~\mathrm{GeV}/c$ and to be reconstructed in $|\eta|<0.8$. To limit the contamination from secondary particles, a selection on the distance of closest approach $(\mathrm{DCA})$ to the reconstructed vertex in the direction parallel to the beam axis $(z)$ of $|\mathrm{DCA}_{z}| < 2\,\mathrm{cm}$ is applied. Also, a \pt-dependent selection on the DCA in the transverse plane $(\mathrm{DCA}_{xy})$ of the selected tracks to the primary vertex is applied. Moreover, tracks associated with the decay products of weakly decaying kaons (``kinks'') are rejected. In ALICE, the set of tracks reconstructed with the above-mentioned selection criteria is commonly referred to as ``global tracks''. 

The use of global tracks yields a significantly non-uniform efficiency as a function of the azimuthal angle and pseudorapidity. In order to obtain a high and uniform tracking efficiency together with good momentum resolution, ``hybrid tracks'' are used~\cite{ALICE:2012nbx,ALICE:2012eyl}. Hybrid tracks correspond to the union of two different sets of tracks selected with complementary criteria: (i) tracks containing at least one space-point reconstructed in one of the two innermost layers of the \ITS (global tracks) and (ii) tracks without an associated hit in the \SPD for which the position of the reconstructed primary vertex is used in the fit of the tracks. Hybrid tracks are used to select the leading particle, as well as to measure \NT and the \pt spectra. Furthermore, in order to select high-quality high-\pt tracks, a selection based on the geometrical track length $(L)$ is applied~\cite{ALICE:2018vuu}. This selection criterion excludes the information from the readout pads at the \TPC sector boundaries ($\approx 3\,\mathrm{cm}$ from the sector edges).

\subsection{Particle identification}
\label{subsec:PID}

ALICE's tracking and particle identification (PID) capabilities allow measuring the transverse momentum spectra of \pion, \kaon, and \pr over a wide range of transverse momentum. In this study the \pt spectra are measured in the $\pt < \ptleading$ interval, using the standard particle identification techniques which have been reported in previous ALICE publications~\cite{RAA_piKp_PbPb_276,pikp_vs_mult_PbPb_highpT_276TeV,Mult_depdencen_piKp_HighpT_5TeV_pPb,pikp_vs_mult_PbPb_5TeV,pikp_vs_mult_13TeV}. Table~\ref{tab:pt_ranges} shows the three techniques used for the PID and the \pt intervals each method covers. 

\begin{table}[!ht]
\centering
\caption{The name of the analysis technique and the transverse momentum ranges in which \pion, \kaon and \pr are identified.}
\begin{tabular}{cccc}
\hline 
Analysis &  \multicolumn{3}{c}{\pt ranges $(\rm{GeV}/c)$} \\
& \pion & \kaon & \pr \\
%\hline
TPC &  $0.25-0.7$ & $0.3-0.6$ & $0.45-1.0$ \\
%\hline
TOF & $0.7-3.0$ & $0.6-3.0$ & $1.0-3.0$ \\
%\hline
rTPC & $2.0-5.0$ & $3.0-5.0$ & $3.0-5.0$ \\
\hline
\end{tabular}
\label{tab:pt_ranges}
\end{table}

At low \pt, the average energy loss, $\langle \dEdx\rangle$, is proportional to $1/(\beta \gamma)^2$ and the relatively large $\pi-\mathrm{K}$ and $\mathrm{p}-\mathrm{K}$ separation power makes it possible to perform  particle identification in this region on a track-by-track basis~\cite{pikp_vs_mult_PbPb_5TeV}. Thus in the \TPC analysis, the relative particle abundances, which are defined as the measured fractions of \pion, \kaon, and \pr with respect to all the measured primary charged particles are obtained from fitting \nsigma distributions in narrow intervals of transverse momentum. For each track, the \nsigma is defined as the difference between the measured and expected \dEdx values normalised to the resolution, $\nsigma = ( \dEdx_{\mathrm{measured}} - \langle \dEdx_{\mathrm{expected}} \rangle ) / \sigma$. 
While the signal of \pion and \pr can be fitted with a Gaussian parameterisation, the one for \kaon uses the sum of two Gaussians as parameterisation to take into account the contamination by electrons. 

In the \TOF analysis, the particle abundances are also measured on a track-by-track basis by fitting the measured $\beta$\footnote{$\beta = L/c\Delta t$, where $L$ is the track length and $\Delta t$ is the measured time-of-flight.} distributions in momentum intervals. In the interval $1 < p < 2~\gevc$, the $\pi-\mathrm{K}$ and $\mathrm{p}-\mathrm{K}$ separation power of hadron identification is large enough~\cite{pikp_vs_mult_PbPb_5TeV} such that one can perform single fits to the signal of \pion, \kaon, and \pr using a Gaussian parameterisation convoluted with an exponential tail. The parameters ($\mu$, $\sigma$ and $\xi$, where $\mu$ and $\sigma$ represent the mean and standard deviation of the Gaussian paramerisation, and $\xi$ represents the $\beta$ value at which the exponential tail begins) of the single fits are extracted from data in $1 < p < 2~\gevc$ and are used to extrapolate to higher momentum values. Finally, the extrapolated functional forms are used to fit the $\beta$ distributions with the sum of three contributions to describe the signals of the three species simultaneously.

In the rTPC analysis, the method described in~\cite{RAA_piKp_PbPb_276,pikp_vs_mult_PbPb_highpT_276TeV,Mult_depdencen_piKp_HighpT_5TeV_pPb} is used. In the relativistic rise region of the TPC $(3 \lesssim \beta \gamma \lesssim 1000)$, the $\meandEdx$ increases as $\mathrm{log}(\beta\gamma)$ and the $\pion-\kaon$ and $\pr-\kaon$ separation power for hadron identification is almost constant~\cite{pikp_vs_mult_PbPb_5TeV}. The knowledge of these two features makes it possible to perform a two-dimensional fit of the correlation between \dEdx and momentum. In order to accomplish this, the first step is to parameterise the Bethe-Bloch and resolution curves in the relativistic rise region. The Bethe-Bloch parameterisation provides the relation between the $\meandEdx$ and $\beta \gamma$, and the parameterised resolution gives the relation between $\sigma_{\dEdx}$ and $\meandEdx$. For the parameterisation, high-purity samples of identified hadrons are used, namely $\pr(\bar{\pr})$ and \pipm from $\lmb(\almb)$ and \kzero decays, respectively, and $\mathrm{e}^{\pm}$ from $\gamma$-conversions~\cite{RAA_piKp_PbPb_276,pikp_vs_mult_PbPb_highpT_276TeV,Mult_depdencen_piKp_HighpT_5TeV_pPb}. Once the Bethe-Bloch and resolution curves are parameterised, they are used to perform the two-dimensional fit. The two-dimensional fit is only used to improve the Bethe-Bloch parameterisation in the transition to the plateau region. Then, the particle ratios are obtained from one-dimensional fits to the \dEdx distributions in momentum intervals using the sum of four Gaussians as a fit function to describe simultaneously the signal of \pion, \kaon, \pr, and $\mathrm{e}$, where the $\mu$ and $\sigma$ of each of the Gaussian distributions are fixed based on the $\meandEdx(\beta \gamma)$ and $\sigma_{\dEdx}(\meandEdx)$ obtained with the above procedure.

\subsection{Corrections}
\label{subsec:corrections_normalization}

The \pt spectra of \pion, \kaon, and \pr are corrected for acceptance and reconstruction inefficiency. The spectra measured with the \TOF detector are also corrected for \TPC--\TOF matching inefficiency. The acceptance and efficiencies are obtained from simulations using the PYTHIA8 Monte Carlo event generator with the Monash 2013 tune (indicated as PYTHIA8 Monash in the following)~\cite{Skands:2014pea}. Subsequently, the propagation of simulated particles through the ALICE apparatus is carried out using GEANT3~\cite{Geant3}. The simulated events are reconstructed using the same algorithms as for the data. The obtained acceptance and reconstruction efficiencies are independent of  the charged-particle multiplicity. Hence, the \RT-integrated values are applied for all the \RT classes. As GEANT3 does not fully describe the interaction of low-momentum $\rm{\overline{p}}$ and $\rm{K^{-}}$ with the detector material, an additional correction factor to the efficiency for these two particles is estimated with GEANT4~\cite{Geant4} and FLUKA~\cite{Fluka}, respectively. These corrections are the same as the ones applied in~\cite{pikp_vs_mult_13TeV}.

The \pt spectra of \pion and \pr contain a large contribution from secondary particles from interactions in the material and particle decays ($\pi^{\pm}$ from $\mathrm{K}^{0}_{\mathrm{S}}$ and $\mathrm{p}(\mathrm{\overline{p}})$ from $\Lambda$ and $\Sigma^{+}$). Since the strangeness production is underestimated in the Monte Carlo event generators, a data-driven approach is used to estimate the fraction of non-primary particles as a function of \pt so that it can be subtracted from the measured spectra. The estimation of this correction is based on a multi-template fit method to describe the measured $\mathrm{DCA}_{xy}$ distributions~\cite{piKp_PbPb_276}. In practice, three Monte Carlo templates representing the expected shapes of $\mathrm{DCA}_{xy}$ distributions of primary particles, secondaries from weak decays, and secondaries from interactions in the material are used to fit the data $\mathrm{DCA}_{xy}$ distributions. The fits are performed in $|\mathrm{DCA}_{xy}| \leq 3~\mathrm{cm}$ and in \pt bins. Since the \TOF analysis only uses tracks matched with the \TOF detector, these corrections are estimated separately for the low- and intermediate-\pt regions. At $\pt = 0.45 ~\gevc$ the contribution from non-primary $\pi^{+} (\rm{p})$ was found to be about $4\% (20\%)$ while at $\pt = 2.0 ~\gevc$ it decreases to about $1\% (4\%)$. Furthermore the correction decreases asymptotically at higher \pt. Therefore, the correction for the \TOF is extrapolated to higher \pt and then applied.

\subsection{Unfolding the charged-particle multiplicity distributions}
\label{subsubsec:one_dimensional_unfolding}

\begin{figure}[!ht]
    \centering
    \hspace{0cm}
    \includegraphics[width=0.48\textwidth]{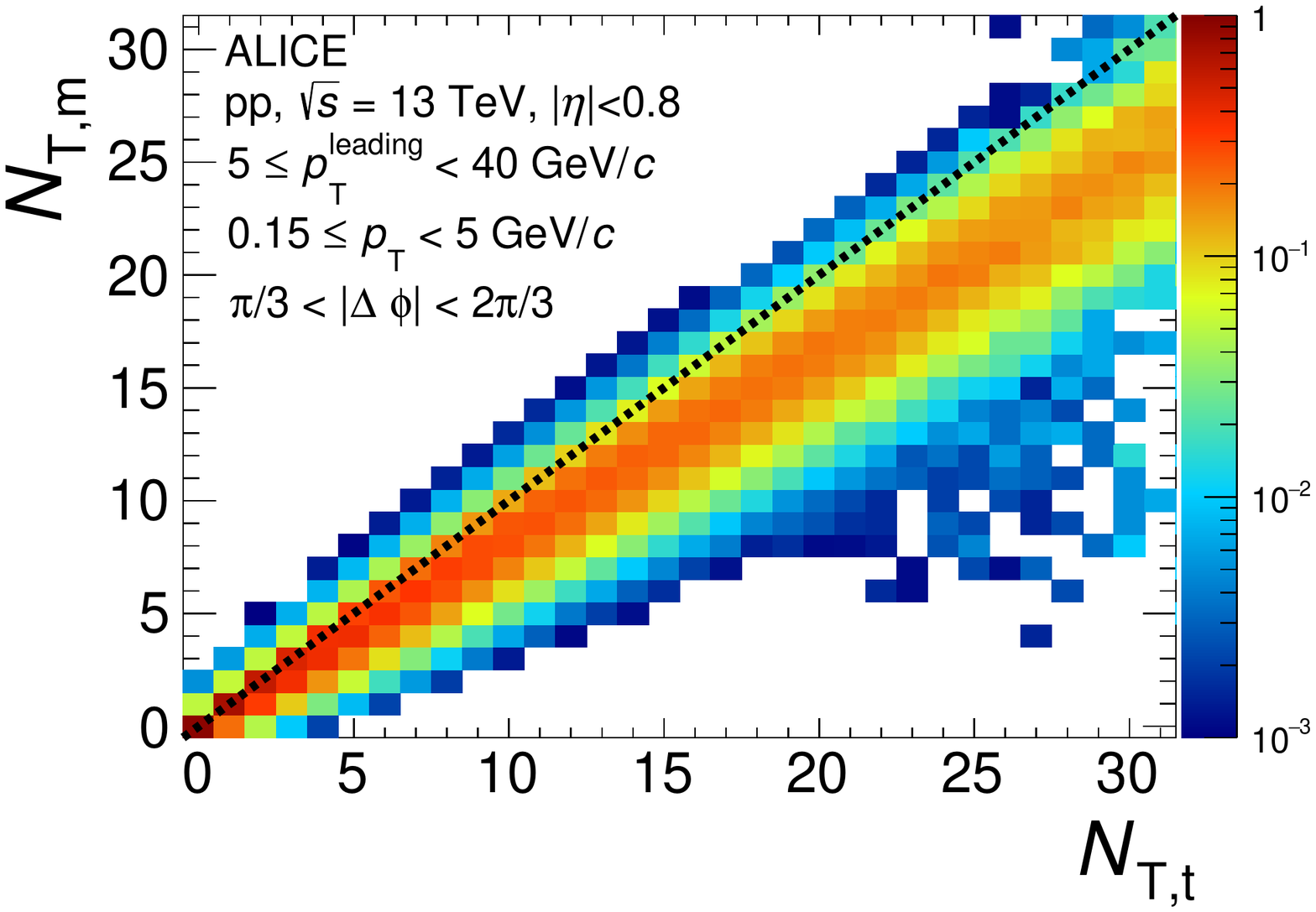}
    \hspace{0cm}
    \includegraphics[width=0.48\textwidth]{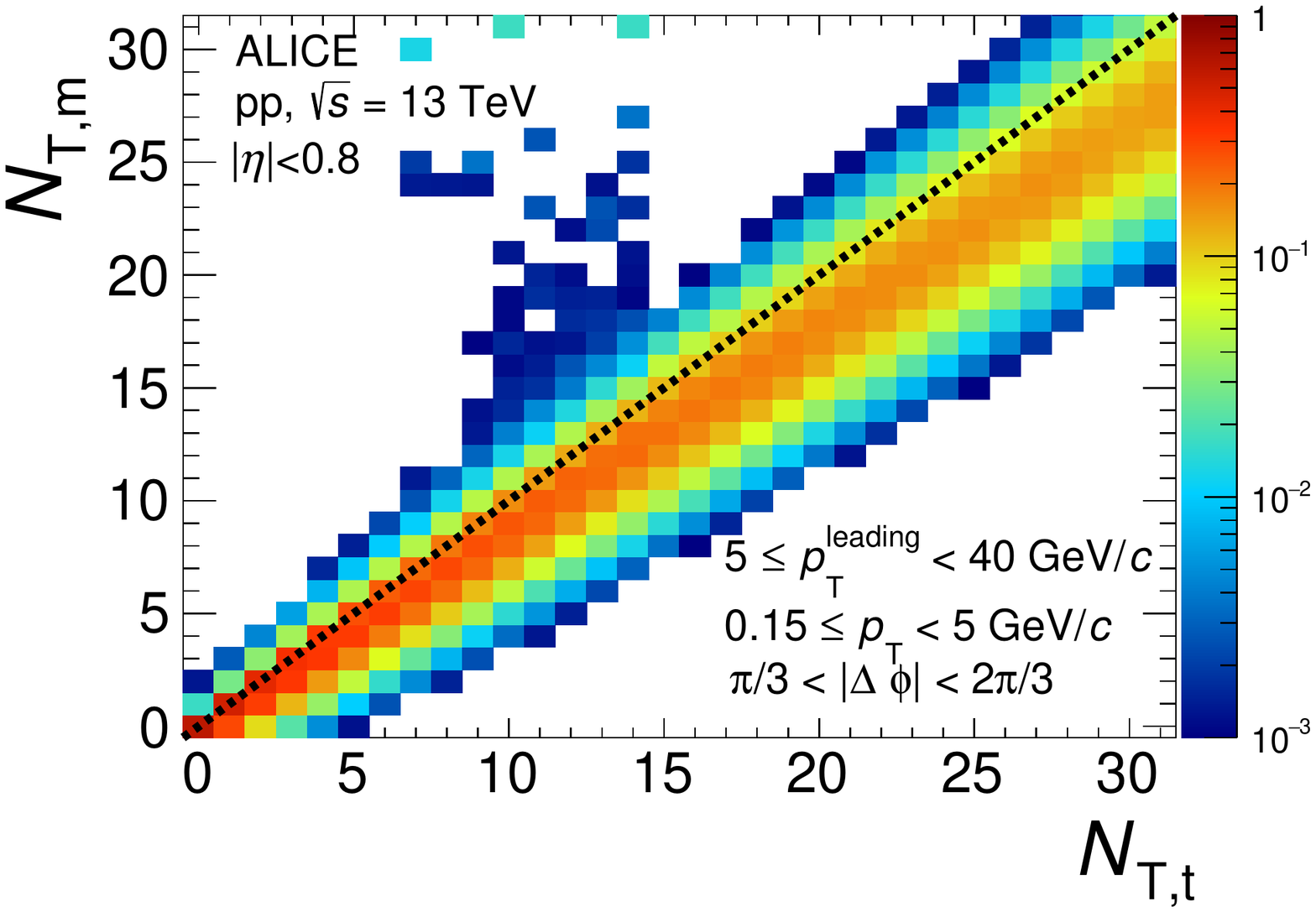}
    \caption{(Left) Correlation between the true \NTt and the measured \NTm multiplicity in the transverse region. (Right) Unfolding matrix \Mone. The iteration step of the unfolding matrix corresponds to the third.}
    \label{fig:RM}
\end{figure}

The charged-particle multiplicity in the transverse region, \NT, is used to characterise the event activity. However, the limited acceptance and finite resolution of the detector cause a smearing of the measured charged-particle multiplicity distribution $Y(\NTm)$. This section introduces the one-dimensional unfolding method to correct for these detector effects and efficiency losses. The adopted approach is based on the iterative Bayesian unfolding method by G. D'Agostini~\cite{DAgostini:1994fjx}. Bayesian unfolding requires the knowledge of the smearing matrix \Smt, which comprises information about the limited acceptance and finite resolution. It represents the conditional probability \RMp of an event with the true multiplicity \NTt to be measured as one with multiplicity \NTm. Figure~\ref{fig:RM}  (left) shows the smearing matrix obtained with simulated events using PYTHIA8 Monash. The values along the diagonal of the smearing matrix represent the probability that a measured event is reconstructed with the true number of particles. At the same time, the off-diagonal elements give the probability that fewer or more particles are reconstructed due to detector inefficiencies and background, e.g., secondary particles misidentified as primary particles.

The one-dimensional unfolded distribution \UnfDis is given as the linear combination between the elements of the unfolding matrix \Mone (see the right panel of Fig.~\ref{fig:RM}) and the measured distribution, 

\begin{equation}
\UnfDis = \sum_{\mathrm{m}} \Mone \RawDis \quad \mathrm{,where} \quad \Mone = \frac{\RMp \, \Prior }{\sum_{t} \RMp \, \Prior}~.
\label{eq:unf_eq}
\end{equation}

\Prior represents a prior probability distribution. It can be any arbitrary distribution at the start of the unfolding process. Here, the measured multiplicity distribution is used as the prior distribution. An updated prior distribution,

\begin{equation}
    \widehat{P}(\NTt) = \frac{Y(\NTt)}{\sum_{\NTt}Y(\NTt)}~,
    \label{eq:UpdatedPrior}
\end{equation}

is obtained from the second iteration and onwards. Thus, the unfolding matrix is improved as the prior distribution is updated. Finally, a new unfolded distribution closer to the true one can be obtained using Eq.~\ref{eq:unf_eq} with the updated \Mone. The smearing in Fig.~\ref{fig:RM} left shows very few events below the main correlation band between $7 < \NTt < 15$ and $15 < \NTm < 30$. This small population comes from statistical fluctuations of the response matrix. Since the unfolding matrix \Mone is proportional to \RMp, these events show up in \Mone in the intervals $22<\NTt< 30$ and $7 < \NTm < 17$, as can be seen in Fig.~~\ref{fig:RM} right. However, given their very small contribution, they are not affecting the unfolding process. 

This iterative process makes the unfolded distribution to converge to the true one eventually. However, it also compounds the effects of statistical uncertainties in the smearing matrix. Therefore, a larger number of iterations does not guarantee a better result: eventually, the true distribution might be contaminated by statistical fluctuations~\cite{Adye:2011gm}. In order to decide when to stop the iterations, the $\chi^{2}/N_{\mathrm{df}}$ between the unfolded and the true distribution as a function of the number of iterations is computed for a Monte Carlo generated sample. The minimum value of the ratio $\chi^{2}/N_{\mathrm{df}}$ indicates when to stop the iterative process. This study found that the optimal number of iterations is three.

\subsection{Unfolding the $\mathbf{\textit{p}_{T}}$ spectra}
\label{subsubsec:unfolding_pt_spectra}

Unfolding the transverse momentum spectra as a function of the multiplicity is treated differently depending on the topological region. The toward and away regions are straightforward cases as there is no overlap between the tracks used for the spectra and the tracks used for the multiplicity calculation as the latter is measured in the transverse region. Therefore, the one-dimensional unfolding matrix \Mone is directly applied  in these two regions. This also makes it trivial to see that the same unfolding matrix can be used for all identified particle spectra. Hence, the fully corrected \pt spectra as a function of \NTt are obtained in a two-step procedure:

\begin{enumerate}
    \item Correct the raw \pt spectra at particle level for tracking inefficiency and secondary particle contamination. The efficiency correction is applied here as the one-dimensional unfolding only affects the classification of the events. 
    
    \item Apply the one-dimensional unfolding matrix. The spectra as a function of \NTt are given by: $\frac{\mathrm{d}\UnfDis}{\mathrm{d}p_{\mathrm{T}}}  = \sum_{\mathrm{m}} \Mone \frac{\mathrm{d}\RawDis}{\mathrm{d}p_{\mathrm{T}}}$ 
\end{enumerate}

The transverse region requires a more elaborate method since both \pt spectra and multiplicity are measured using the same tracks. In other words, one is no longer dealing with the problem of rearranging events but rather how tracks should be unshuffled to match the true transverse momentum distributions. This poses a multi-dimensional problem with two dimensions associated to the true and measured multiplicities and two additional dimensions (true and measured yields) for each \pt bin. Instead of performing the full multi-dimensional unfolding, an approximate method is employed in which the multiplicity smearing matrix is assumed to be independent of the transverse momentum. This is a very good approximation as the efficiency is essentially flat in \pt for the track selection and \pt ranges used here. In this approach, a new response matrix is obtained by multiplying every column of the original multiplicity response matrix with the respective number of measured particles as weights. After row-wise normalisation, the desired track smearing matrix is obtained.

The unfolding is done bin-by-bin in \pt with this modified response matrix. For a particular transverse momentum bin, the measured multiplicity distribution is unfolded using the iterative unfolding procedure described in Sec.~\ref{subsubsec:one_dimensional_unfolding}. This approach yields unfolding matrices that depend on the transverse momentum. Henceforth, these matrices will be called \Mtwo. It should be stressed that this method works here because the tracking efficiency does not depend strongly on the transverse momentum for hybrid tracks and because the same tracks to measure \NT are used to obtain the spectra. 

Similar to the toward and away regions, the two-step procedure is followed to obtain the fully corrected transverse momentum spectra. The only difference is that in the transverse region the \pt-dependent \Mtwo matrices are used

\begin{equation}
    \frac{\mathrm{d}Y(\NTt,\pt)}{\mathrm{d}p_{\mathrm{T}}}  = \sum_{\mathrm{m}} \Mtwo \frac{\mathrm{d}Y(\NTm,\pt)}{\mathrm{d}p_{\mathrm{T}}}.
    \label{eq:unf_tr}
\end{equation}

The method described above unfolds the spectra of all charged particles and yields the unfolding matrices \Mone and \Mtwo. When unfolding the spectra of identified particles (for example, \pion in the transverse region), Eq.~\ref{eq:unf_tr} is applied using the \Mtwo matrices from charged particles and then exchanging $\textrm{d}Y(\NTm,\pt)/\textrm{d}\pt$ for $\textrm{d}Y^{\pion}(\NTm,\pt)/\textrm{d}\pt$. The unfolding of \pion spectra in the toward and away regions is done with the same strategy but using \Mone instead.

\section{Systematic uncertainties}
\label{sec:systematic_uncertainties}

In this section, the estimation of the systematic uncertainties is described. The systematic uncertainties on the \pt spectra are divided into two categories, \RT-dependent and \RT-independent uncertainties. The total systematic uncertainty on the \pt spectra is given as the sum in quadrature of all the individual sources of uncertainty.

\textbf{\RT-dependent systematic uncertainties}

The unfolding method described in Sec.~\ref{subsubsec:one_dimensional_unfolding} shows deficiencies, mainly when unfolding the \pt spectra for low multiplicities in the transverse region. To account for these deficiencies, the following contributions to the systematic uncertainty on the \NT distribution are considered:

\begin{itemize}
    \item Monte Carlo (MC) non-closure: PYTHIA8 Monash is the default tune for the generation of the multiplicity response matrix and \NT distributions with and without the detector's efficiency losses. The unfolded \NT spectrum from the simulation is compared to the generated one. Thus, any statistically significant difference between the generated and unfolded distributions is referred to as MC non-closure and is added in quadrature to the total systematic uncertainty. During the unfolding procedure, the MC closure improves with the number of iterations, with an optimal number of three, which leads to a negligible MC non-closure.
    
    \item Dependence on the choice of the MC model: EPOS LHC~\cite{EPOS_LHC} is used to generate a different multiplicity response matrix. This response matrix is used to unfold the \NT and \pt spectra. The ratio between the final unfolded distributions using PYTHIA8 Monash and EPOS LHC was quantified and added to the total systematic uncertainty. In the interval $0 < \NT < 18$, the relative systematic uncertainty is below $2\,\%$, increasing to about $4\,\%$ at $\NT \approx 18$. Due to statistical limitations on the response matrix, a constant $4\,\%$ relative systematic uncertainty for $\NT \geq 18$ was assigned.
    
    \item Track selection: This uncertainty is quantified by changing the track selection criteria with respect to the nominal one. In particular, the minimum number of crossed rows in the \TPC is set to $60$ and $100$ (the nominal is $70$). The track fit quality in the \ITS and \TPC quantified by the $\chi^{2}_{\ITS}/N_{\mathrm{hits}}$ and the $\chi^{2}_{\TPC}/N_{\mathrm{clusters}}$ must not exceed $25$ and $49$ (the nominal is $36$), and $3$ and $5$ (the nominal is $4$), respectively. The maximum distance of closest approach to the vertex along the beam axis $(\mathrm{DCA}_{z})$ is set to $1$ and $5\,\mathrm{cm}$ (the nominal is $2\,\mathrm{cm}$). Furthermore, the parameters of the geometrical length cut to select the leading particle are also varied. For a particular parameter variation, the maximum difference between the results obtained with the tighter and looser selections with respect to the nominal value is quantified. The total systematic uncertainty from track variations is given as the sum in quadrature of the different parameter variations. The relative systematic uncertainty is on average $1\,\%$ in the interval $0 < \NT < 18$ and increases for higher \NT values. For $\NT \geq 18$, the statistical fluctuations become significant. Therefore, a constant $2\,\%$ relative systematic uncertainty was assigned.      
\end{itemize}

\textbf{\RT-independent systematic uncertainties}

The \RT-independent systematic uncertainties are divided into two categories. The first category includes the uncertainties common to the different analyses, such as those due to the track quality criteria and the \pt-dependent \ITS--\TPC matching efficiency. The \ITS--\TPC matching efficiency is derived from matching \ITS pure tracks with the corresponding ITS+TPC tracks (in the same phase-space region) and by comparing the matching efficiency in data and Monte Carlo simulations. The second category groups the analysis specific uncertainties. It includes the uncertainties on the secondary particle contamination correction estimation, the signal extraction technique and the \TPC--\TOF matching efficiency.

As described in Sec.~\ref{subsec:corrections_normalization}, the secondary particle contamination correction is based on multi-template fits to the $\mathrm{DCA}_{xy}$ distributions in transverse momentum intervals. The estimation of the systematic uncertainty follows the procedure described in~\cite{pikp_vs_mult_PbPb_5TeV}. Namely, the fitting range is changed from the nominal values of $\pm 3\,\mathrm{cm}$ to $\pm 1.5\,\mathrm{cm}$.

To estimate possible systematic effects attributed to the signal extraction technique in the \TPC analysis, a similar procedure to the one described in~\cite{pikp_vs_mult_PbPb_5TeV} was applied. The signal extraction technique changed from fitting \nsigma distributions to bin counting in the range of $\pm 3\sigma$. The systematic uncertainty on the particle fractions is given as the difference between the nominal particle fractions and the ones obtained from bin counting.

As described in Sec~\ref{subsec:PID}, the measurement of the particle fractions in the \TOF analysis is based on fits to $\beta$ distributions in momentum intervals. Hence, the systematic uncertainty is mainly driven by the uncertainty in the parameterisation of the $\mu, \sigma$, and $\xi$ curves for \pion, \kaon, and \pr. The relative difference between the fitted curves and the actual
measured $\mu, \sigma$, and $\xi$ values was computed to evaluate the effect of the parameterisations. Thus, the systematic uncertainty in the extraction of the particle fractions is obtained by refitting the $\beta$ distributions while randomly varying the constrained parameters $\mu, \sigma$, and $\xi$ within the uncertainty of the parameterisations assuming a Gaussian variation centred at the nominal value. The refitting was performed 1000 times, and the systematic uncertainty on the particle fractions as a function of the transverse momentum is given as the standard deviation of the associated distributions. This approach is motivated by work developed in~\cite{RAA_piKp_PbPb_276,Mult_depdencen_piKp_HighpT_5TeV_pPb,pikp_vs_mult_PbPb_5TeV}.

The measurement of the systematic uncertainty on the extraction of the particle fractions in the rTPC analysis follows the method from~\cite{RAA_piKp_PbPb_276, pikp_vs_mult_PbPb_5TeV,Mult_depdencen_piKp_HighpT_5TeV_pPb}. In this analysis, the primary source of systematic uncertainty comes from the imprecise description of the detector response, namely the Bethe-Bloch and resolution parameterisations. To estimate the systematic effect, the relative difference between the parameterisations and the actual $\langle \dEdx \rangle$ and $\sigma_{\dEdx}$ values are measured. The particle fractions are measured following a fitting procedure where the constrained parameters, $\langle \dEdx \rangle$ and $\sigma_{\dEdx}$, are allowed to vary randomly within the uncertainty of the parameterisations. The fitting procedure was repeated 1000 times and the systematic uncertainty in the particle fractions is given as the standard deviation of the associated distributions.

When computing the \pt-differential particle ratios, all the systematic uncertainties cancel out in the ratios except those attributed to the signal extraction and feed-down. In the high \pt region (rTPC analysis) the procedure described in~\cite{RAA_piKp_PbPb_276} is used to extract the signal extraction systematic uncertainty on the \ktopi and \ptopi ratios directly from fits to the \dEdx distributions.

Table~\ref{tab:syspikp} lists a summary of the systematic uncertainties at different \pt values for the spectra and particle ratios in the transverse region. The table is divided into common and analysis-specific uncertainties. The values in the toward and away regions are the same as those of the transverse region. The only topological-region-dependent uncertainty is the one attributed to the MC non-closure.

\begin{table*}[ph!]\label{syst}
\caption{Summary of systematic uncertainties on the \pion,  \kaon, and \pr \pt spectra. The uncertainties are shown for different representative \pt values. The last two rows show the total systematic uncertainty on the \pt spectra and the \pt-differential particle ratios. These values correspond to the spectra in the transverse region in the $0 \leq \RT < 0.5$ class.}
\begin{tabularx}{\textwidth}{p{5.3cm}*{2}{Y}*{1}{Y | }*{2}{Y}*{1}{Y | }*{3}{Y}}
\hline\hline
    &  \multicolumn{9}{c}{Uncertainty (\%)} \\
    {\bf Common source}  & \multicolumn{3}{c}{\pion} & \multicolumn{3}{c}{\kaon} & \multicolumn{3}{c}{\pr} \\
\hline
    \pt\ (\GeVc)   &   0.3 & 2 & 5   &   0.3 & 2 & 5 &   0.45 & 2 & 5 \\
    \hline
    
    \ITS--\TPC matching efficiency  & 1.4 & 2.6 & 2.9 & 1.4 & 2.6 & 2.9 & 1.4 & 2.6 & 2.9\\

    MC non-closure &  & 3.2 & & & 3.6 &  & & 1.5 & \\

    MC dependence & 1 & 1.5 & 1.7 & 0.9 & 1.5 & 1.7 & 0.9 & 1.5 & 2 \\ 

    Track selection &  & 1 &  &  & 1 &  &  & 1 & \\ 

\hline
    {\bf Analysis-specific} & \multicolumn{3}{c}{\pion} & \multicolumn{3}{c}{\kaon} & \multicolumn{3}{c}{\pr} \\
\hline

    {\bf TPC}, \pt\ (\GeVc) & 0.3 & & 0.7 & 0.3 & & 0.6 & 0.45 & & 1\\
    \hline
    PID & 0.1 & & 1.8 & 7.3 & & 5.9 & 0.1 & & 3.4 \\
    Feed-Down & 1 & & 0.3 & - & & - & 10 & & 1.1 \\

\hline\hline

    {\bf TOF}, \pt\ (\GeVc)  & 1 &  & 2   &   1 &  & 2  &   1 &  & 2    \\
    \hline
    PID & negl. & & 1 & 0.3 & & 3.4 & 0.2 & & 0.7 \\
    Feed-Down & 0.3 & & negl. & - & & - & 1 & & 0.2 \\
    TOF matching efficiency & 3 & & 3 & 6 & & 6 & 4 & & 4 \\

\hline\hline

    {\bf rTPC}, \pt\ (\GeVc) & 3 & & 5 & 3 & & 5  &   3 & & 5    \\
    \hline
    PID & 0.7 &  & 0.6 & 6.4 &  & 2.8  & 5.8 & & 4.2 \\
    Feed-Down & negl. & & negl. & - &  & - & 0.2 & & 0.2\\

\hline\hline

   {\bf Total}  & \multicolumn{3}{c}{\pion} & \multicolumn{3}{c}{\kaon} & \multicolumn{3}{c}{\pr} \\
\hline
    \pt\ (\GeVc)   &   0.3 & 2 & 5   &   0.3 & 2 & 5 &   0.45 & 2 & 5   \\
    \hline
    Total & 3.9 & 5.5 & 4.7 & 8.3 & 8.3 & 5.7 & 10.2 & 5.3 & 5.7 \\
\hline\hline
    {\bf Particle ratios}  & \multicolumn{3}{c}{} & \multicolumn{3}{c}{\ktopi} & \multicolumn{3}{c}{\ptopi} \\
\hline
    \pt\ (\GeVc)   &    &  &    &   0.3 & 2 & 5 &   0.45 & 2 & 5 \\
\hline
    Total & & & & 7.4 & 4.1 & 3.2 & 10.1 & 1.5 & 4 \\
    \bottomrule
  \end{tabularx}
  ~\newline
  \label{tab:syspikp}
\end{table*}

\clearpage
\section{Results}\label{sec:results}

This section presents the results of the production of \pion, \kaon, and \pr as a function of the relative transverse activity classifier, \RT. The data are compared with predictions from PYTHIA8 Monash~\cite{Skands:2014pea}, PYTHIA8 with ropes hadronisation model (indicated as PYTHIA8 ropes)~\cite{Bierlich:2014xba}, HERWIG7~\cite{Bahr:2008pv,Bellm:2015jjp}, and EPOS LHC~\cite{EPOS_LHC}. PYTHIA8 with Monash tune is one of the most popular event generators at LHC energies for most observables but lacks the QGP-like effects observed in small collision systems such as strangeness enhancement, while the other three models are known to describe the strangeness enhancement in small collision systems better~\cite{ALICE:2016fzo,Bierlich:2014xba,Gieseke:2017clv}. Hence, these models allow for testing a broad range of possible dynamic effects. In PYTHIA8 Monash, the soft-inclusive particle production is based on multiple perturbative parton--parton interactions (MPI)~\cite{PhysRevD.36.2019}. This model also includes a colour reconnection (CR) mechanism~\cite{Sjostrand:2006za}, allowing each MPI system's partons to be colour connected with a higher-\pt MPI system. In particular, PYTHIA8 Monash describes the enhanced \pt-differential proton-to-pion ratio at intermediate \pt~\cite{pikp_vs_mult_13TeV} by introducing the colour reconnection mechanism and does not need to assume the formation of a medium~\cite{PhysRevLett.111.042001}. PYTHIA8 ropes model allows strings to fuse in an environment with a high density of strings and form ``colour ropes''. Consequently, colour ropes are expected to produce more strange hadrons and baryons, the latter via probabilistic collapses of ropes to string junctions. EPOS LHC is a core-corona model, which assumes the formation of a QGP medium in the high-density core regions in \pp collisions. The hadronisation of the corona is based on string fragmentation, while the particles associated with the core are thermally produced (grand-canonical thermal description). In EPOS LHC, particle production in low-multiplicity events is mainly dominated by string fragmentation. In contrast, high-multiplicity events are core dominated, and a large production of strange hadrons and baryons is expected. Particle production in the HERWIG7 is based on cluster hadronisation and it has its own colour reconnection mechanism where baryonic clusters are allowed to be produced in a geometric manner. This model also includes a non-perturbative gluon splitting mechanism to create more $\mathrm{s}\overline{\mathrm{s}}$ pairs to account for the strangeness enhancement~\cite{Duncan:2018gfk}.   

The \pt spectra as a function of \RT are normalised to the total number of events in each \RT class. The relation between \RT intervals and \NT classes is given in Table~\ref{tab:NumberOfEventsPerRTBin}. The \RT distribution is constructed using the unfolded \NT distribution for which the $\langle \NT \rangle$ is equal to $7.366 \pm 0.002$ (stat.). For each $R_{\mathrm{T}}$ bin the intervals under the $N_{\mathrm{T}}$ column are inclusive meaning that for $0 \leq R_{\mathrm{T}} < 0.5$, $N_{\mathrm{T}}$ is equal to 0, 1, 2 or 3.

\begin{table}[htb]
    \centering
    \caption{Relation between \RT intervals and \NT classes.}
    \begin{tabular}{ccc}
    \hline
        $\RT=\NT/\langle \NT \rangle$ & \NT & Number of events \\
        0--0.5 & 0--3 & 2613151  \\
        0.5--1.5 & 4--11 & 4055410  \\
        1.5--2.5 & 12--18 & 1302116 \\
        2.5--5 & 19--30 & 180652 \\
        0--5 & 0--30 & 8151331 \\
    \hline
    \end{tabular}
    \label{tab:NumberOfEventsPerRTBin}
\end{table}

\begin{figure}[htb]
    \centering
    \hspace{0cm}
    \includegraphics[width=0.48\textwidth]{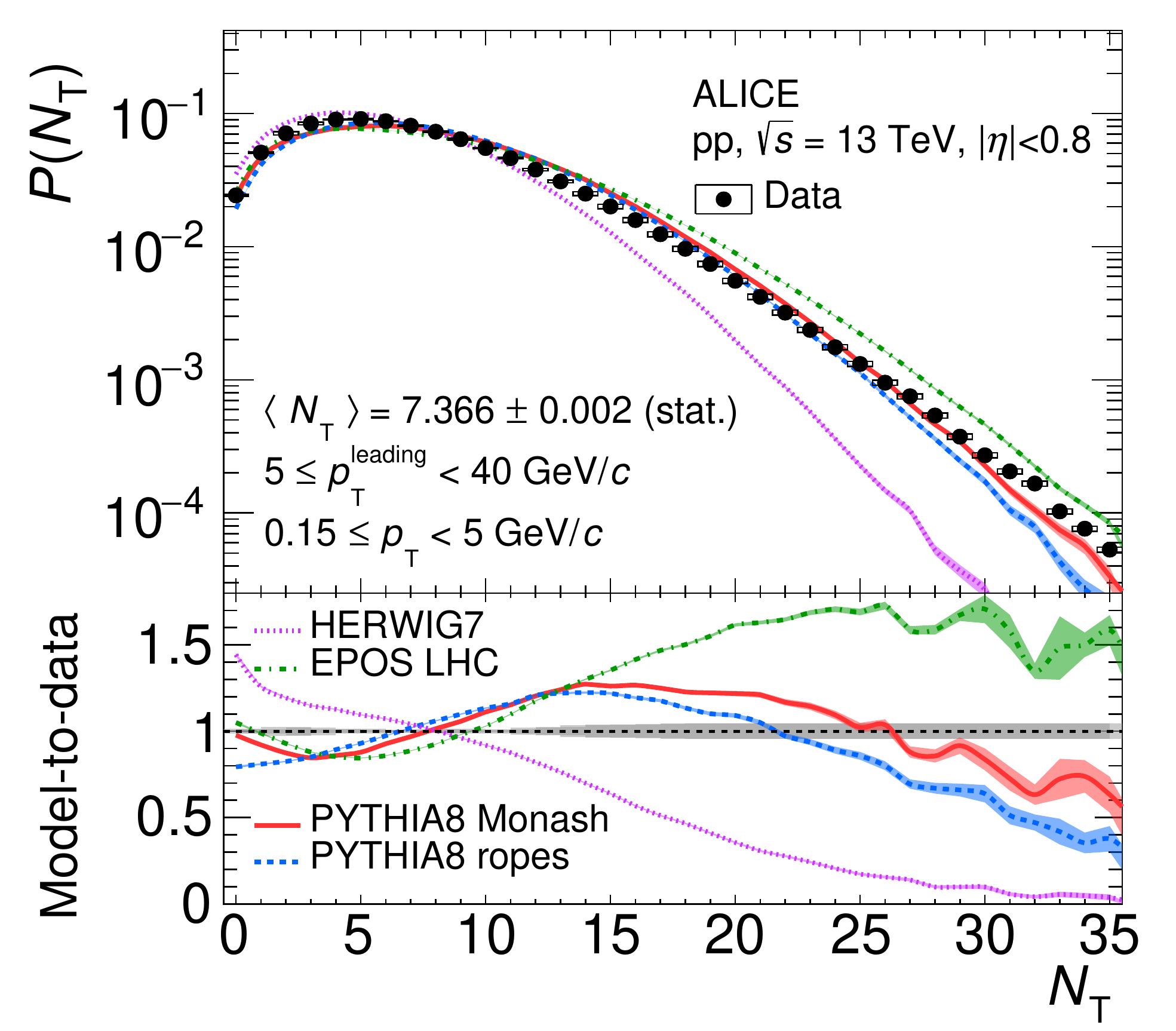}
    \hspace{0cm}
    \includegraphics[width=0.48\textwidth]{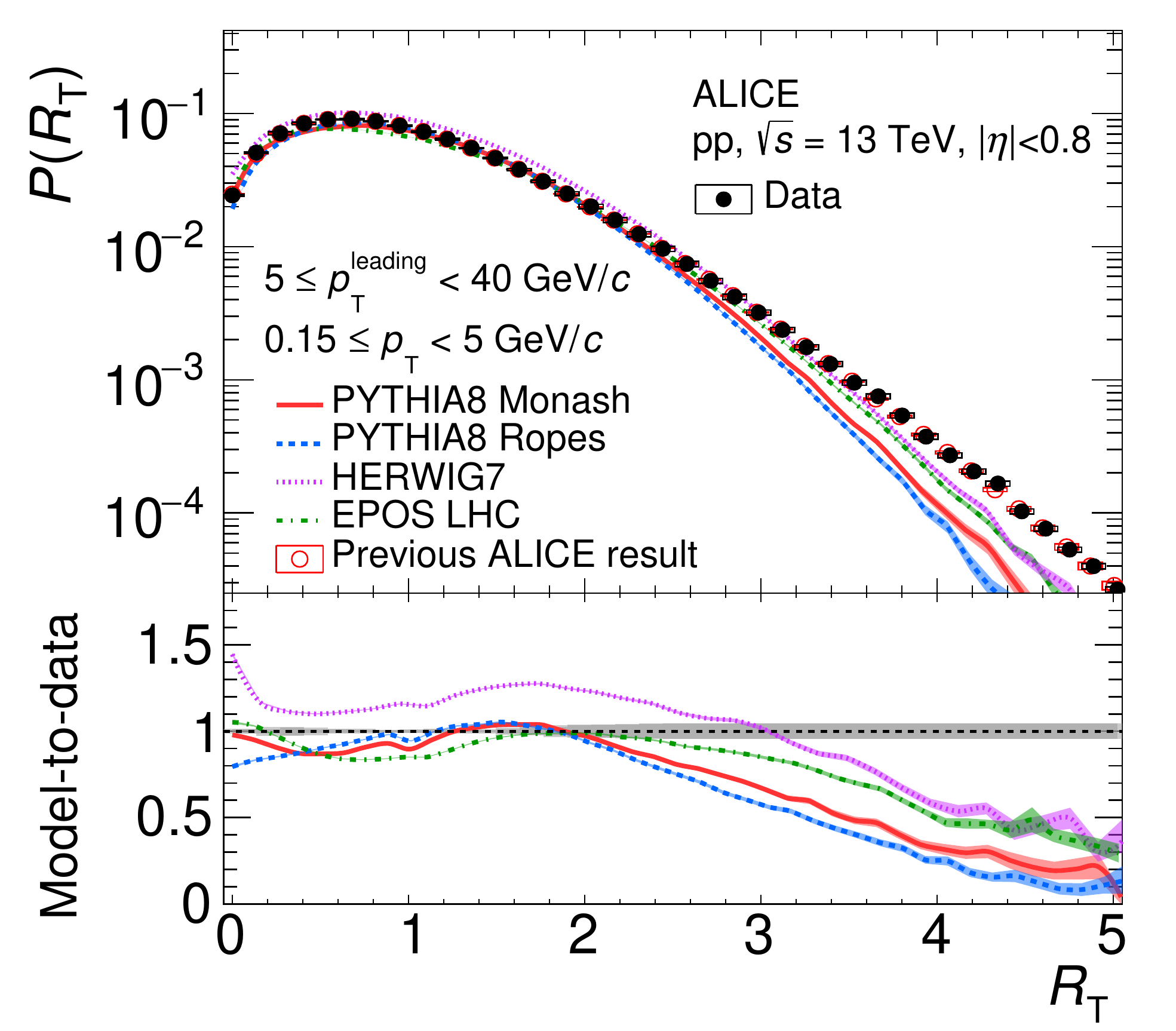}
    \caption{\NT (left) and \RT (right) probability distributions in the transverse region in events with the leading particle. The data are represented with solid black markers and statistical and systematic uncertainties with error bars and boxes, respectively. Model predictions are presented with colour lines and the bands around the model predictions represent only the statistical uncertainty. The bottom panels show the model-to-data ratios. The grey band centred at one in the bottom panel represents the systematic uncertainties of the data.}
    \label{fig:RT}
\end{figure}

Figure~\ref{fig:RT} shows the unfolded \NT and \RT probability distributions in the transverse region integrated over all the events with the leading particle along with different model predictions. For each model, the $\langle \NT \rangle$ corresponds to the mean value of the corresponding \NT spectrum. It is observed that PYTHIA8 Monash and PYTHIA8 ropes give the best qualitative description of the \NT distribution, while EPOS LHC (HERWIG7) overestimates (underestimates) the data for $\NT >10$. However, when \RT is computed, all the models underestimate the data for $\RT \gtrsim 2$. This is because the models poorly describe the low-\NT region, so they predict larger $\langle \NT \rangle$ values than the measured ones. Finally, the \RT probability distribution is compared with the previous ALICE result~\cite{ALICE:2019mmy}, which used a limited data sample and applied the unfolding at the level of the \RT distribution while in the current analysis the \RT spectrum is derived from the \NT distribution. The new result is in agreement with the previous ALICE measurement within 1.5\%.

\begin{figure}[htb]
    \centering
    \hspace{0cm}
    \includegraphics[width=1\textwidth]{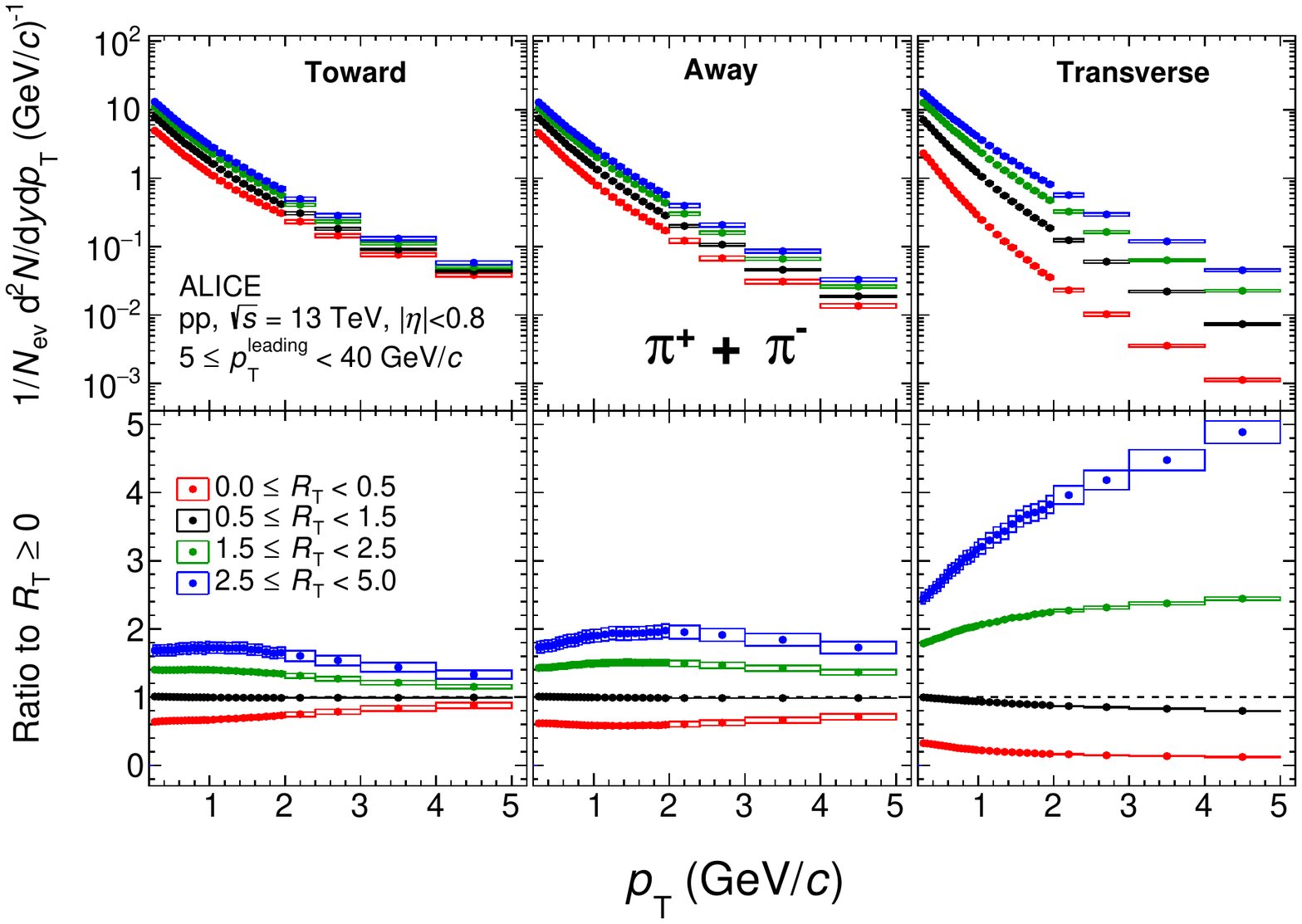}
    \caption{Transverse momentum spectra (top panels) of pions as a function of \RT and ratios to the \RT-integrated spectrum (bottom panels). The toward, away, and transverse regions are shown from left to right. The statistical and systematic uncertainties are represented with bars and boxes, respectively.}
    \label{fig:Spectra_data_pion}
\end{figure}

\begin{figure}[htb]
   \centerline{
     \includegraphics[width=1\columnwidth]{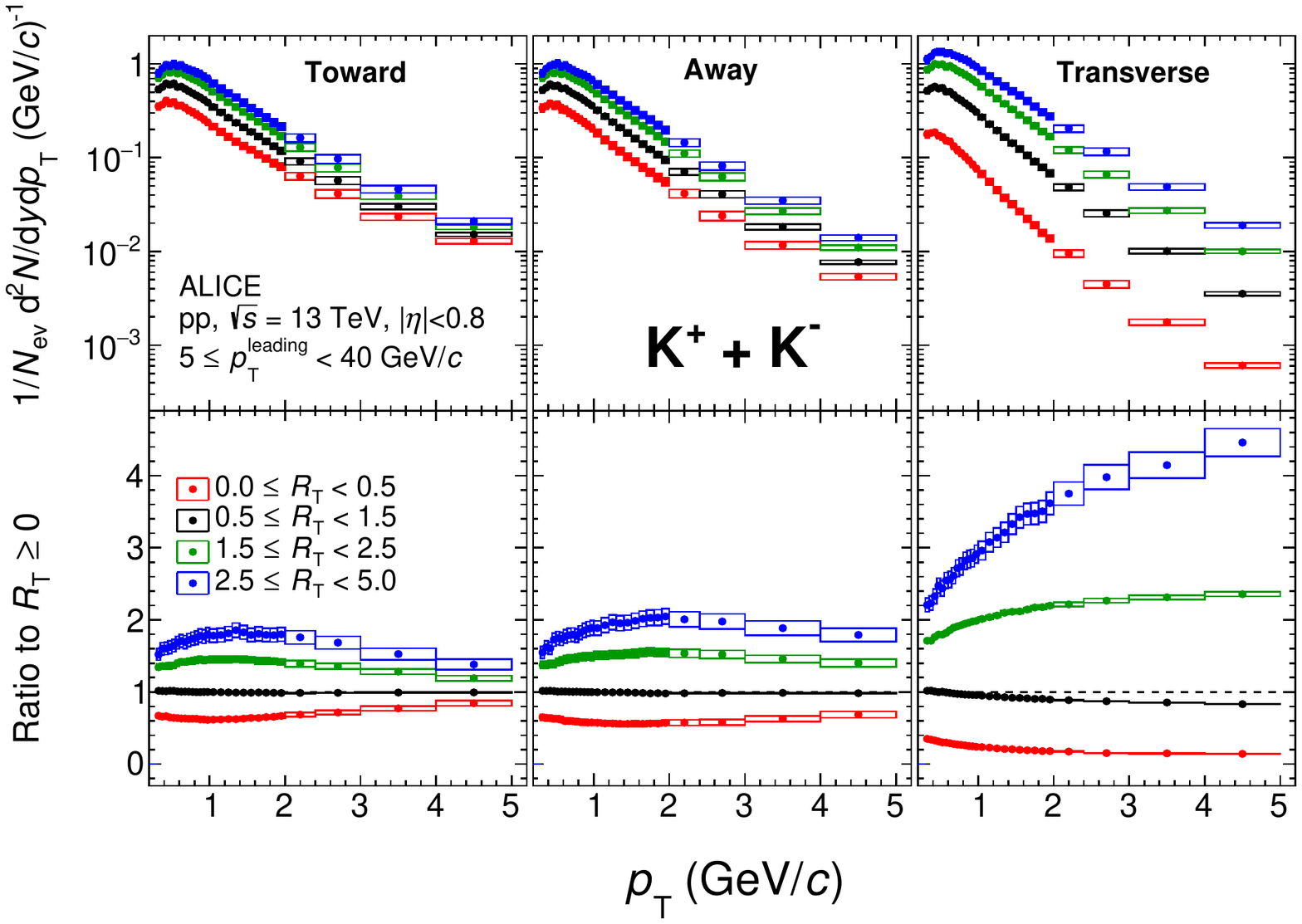}}
   \caption{Transverse momentum spectra (top panels) of kaons as a function of \RT and ratios to the \RT-integrated spectrum (bottom panels). The toward, away, and transverse regions are shown from left to right. The statistical and systematic uncertainties are represented with bars and boxes, respectively.}
   \label{fig:Spectra_data_kaon}
 \end{figure}

\begin{figure}[htb]
    \centering 
    \hspace{0cm}
    \includegraphics[width=1\textwidth]{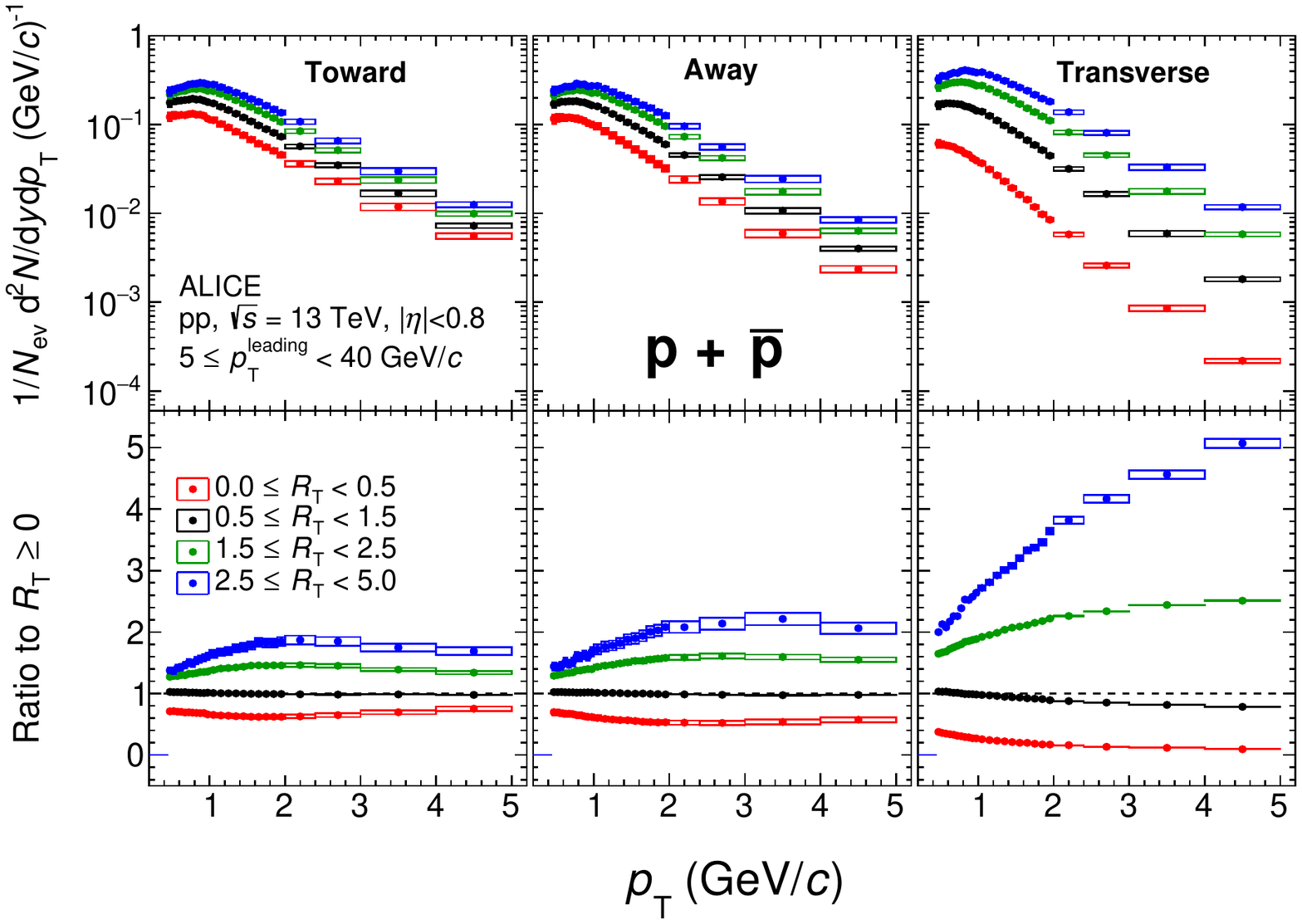}
    \caption{Transverse momentum spectra (top panels) of protons as a function of \RT and ratios to the \RT-integrated spectrum (bottom panels). The toward, away, and transverse regions are shown from left to right. The statistical and systematic uncertainties are represented with bars and boxes, respectively.}
    \label{fig:Spectra_data_proton}
\end{figure}

\begin{figure}[!ht]
    \centering
    \hspace{0cm}
    \includegraphics[width=1\textwidth]{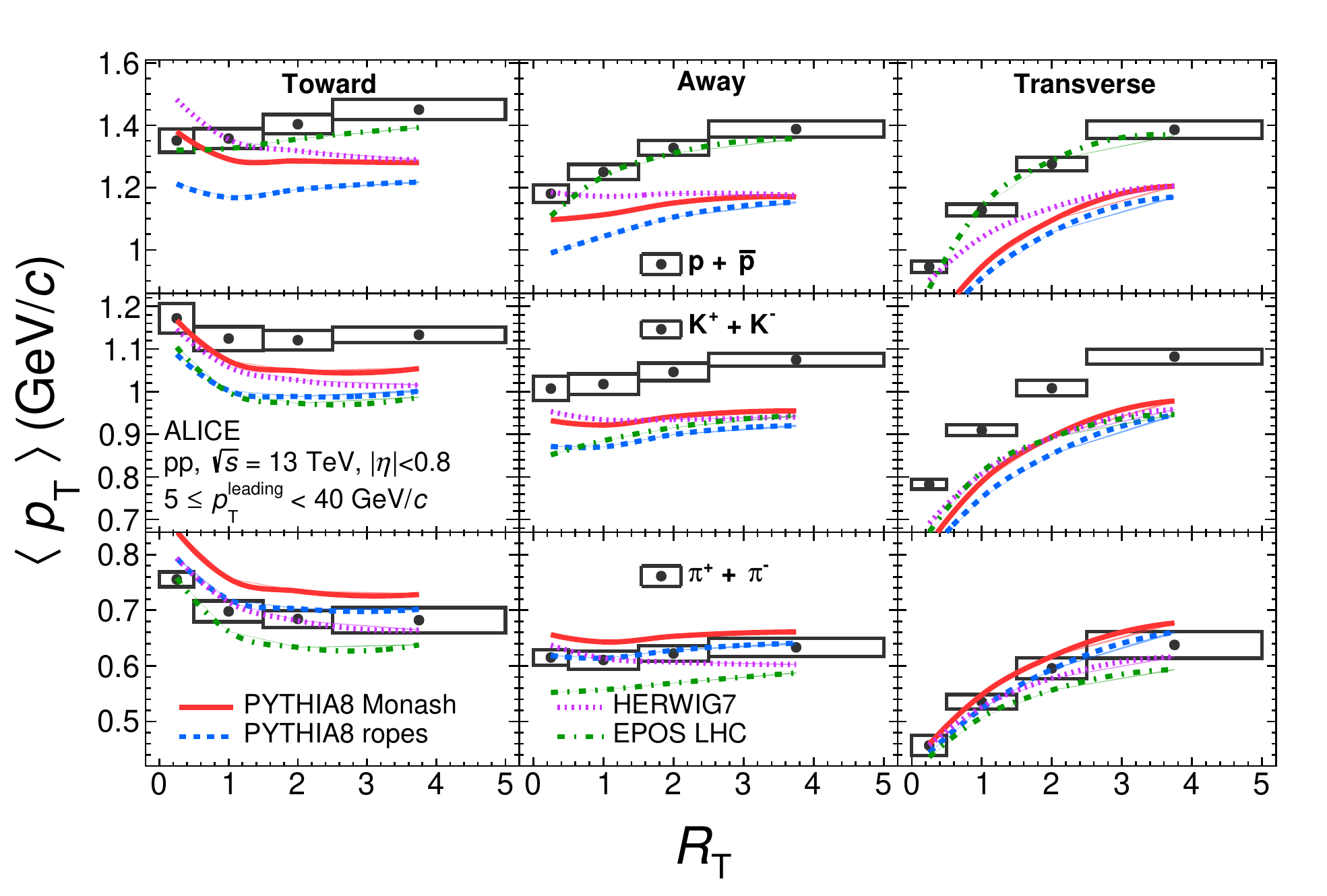}
    \caption{Average transverse momentum as a function of \RT. The \meanpt in the toward, away, and transverse regions are shown from left to right. The results for pion, kaon, and protons are shown in the first, second, and third row, respectively. Statistical and systematic uncertainties are represented with error bars and boxes, respectively. The shaded bands around the model lines represent the statistical uncertainties.}
    \label{fig:MeanpT_wMCs}
\end{figure}

\Cref{fig:Spectra_data_pion,fig:Spectra_data_kaon,fig:Spectra_data_proton} show the transverse momentum distributions of \pion, \kaon, and \pr as a function of \RT. The results in the toward, away, and transverse regions are shown on the left, middle, and right panels, respectively. The lower panels show the ratios between the \RT-dependent \pt spectra and the \RT-integrated \pt spectrum. The \RT-independent systematic uncertainties cancel out in the ratios. The \RT-dependent systematic uncertainties are correlated and cancel out only partly. From the ratios to the \RT-integrated spectrum, it is observed that the toward and away regions share a similar feature at low transverse momentum: a depletion of low-\pt particles with increasing \RT. Furthermore, this effect follows a mass ordering, being larger for heavier particles. This behaviour is reminiscent of radial flow effects, in which the depletion of low-\pt particles is compensated by an increasing number of particles at intermediate \pt. The particle production in the toward and away regions is dominated by the leading and away-side jet fragmentation into high-\pt particles. This can be observed in the ratio between the spectra in $0\leq \RT < 0.5$ and the \RT-integrated ones (bottom panels of~\Cref{fig:Spectra_data_pion,fig:Spectra_data_kaon,fig:Spectra_data_proton}), which increases with \pt (in the interval $\pt \gtrsim 2~\gevc$), and the effect is more evident for pions. The opposite is observed for the spectral shapes at high \RT; they soften with increasing \RT for $\pt \gtrsim 2~\gevc$. This can be interpreted as a ``dilution'' of the jet with increasing UE activity. When $R_{\mathrm{T}} \rightarrow \infty$ the particle multiplicity from the UE is higher than the particle multiplicity from the jet in the toward region. Thus, average quantities like $\langle p_{\mathrm{T}} \rangle$ of pions and kaons in the toward region decreases at high $R_{\mathrm{T}}$ (see Fig.~\ref{fig:MeanpT_wMCs}). The $\langle p_{\mathrm{T}} \rangle$ of protons increases instead with increasing $R_{\mathrm{T}}$ because there other effects like radial flow are more relevant. This can also be seen in the ratios to $R_{\mathrm{T}}$-integrated spectrum, where they decrease with increasing $p_{\mathrm{T}}$ for events with high UE activity. The spectral shapes of all the species in the transverse region share a common feature: they harden with increasing UE activity. This effect can be attributed to jet hardening with increasing multiplicity.

\begin{figure}[htb]
    \centering 
    \hspace{0cm}
    \includegraphics[width=1\textwidth]{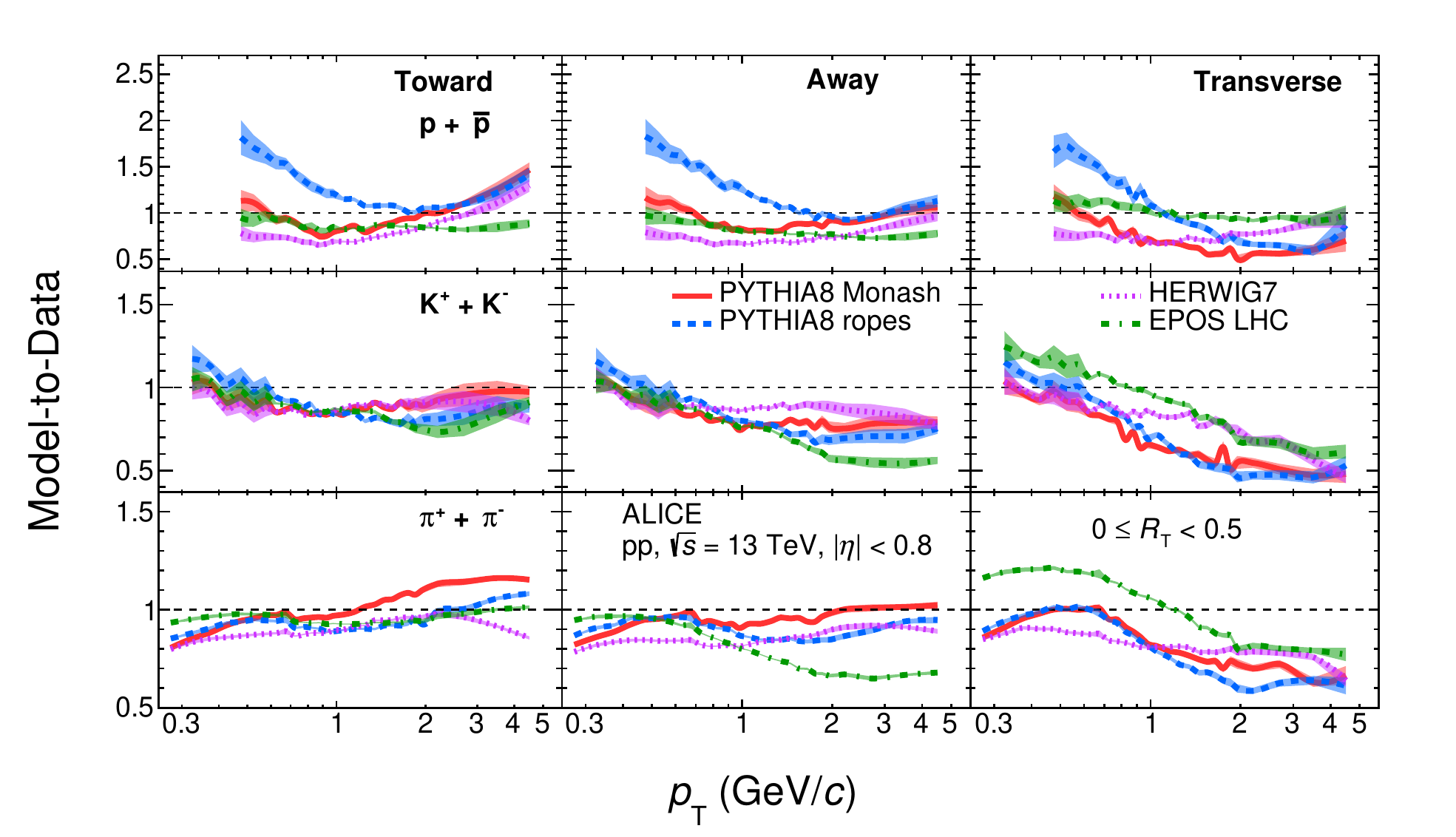}
    \hspace{0cm}
    \includegraphics[width=1\textwidth]{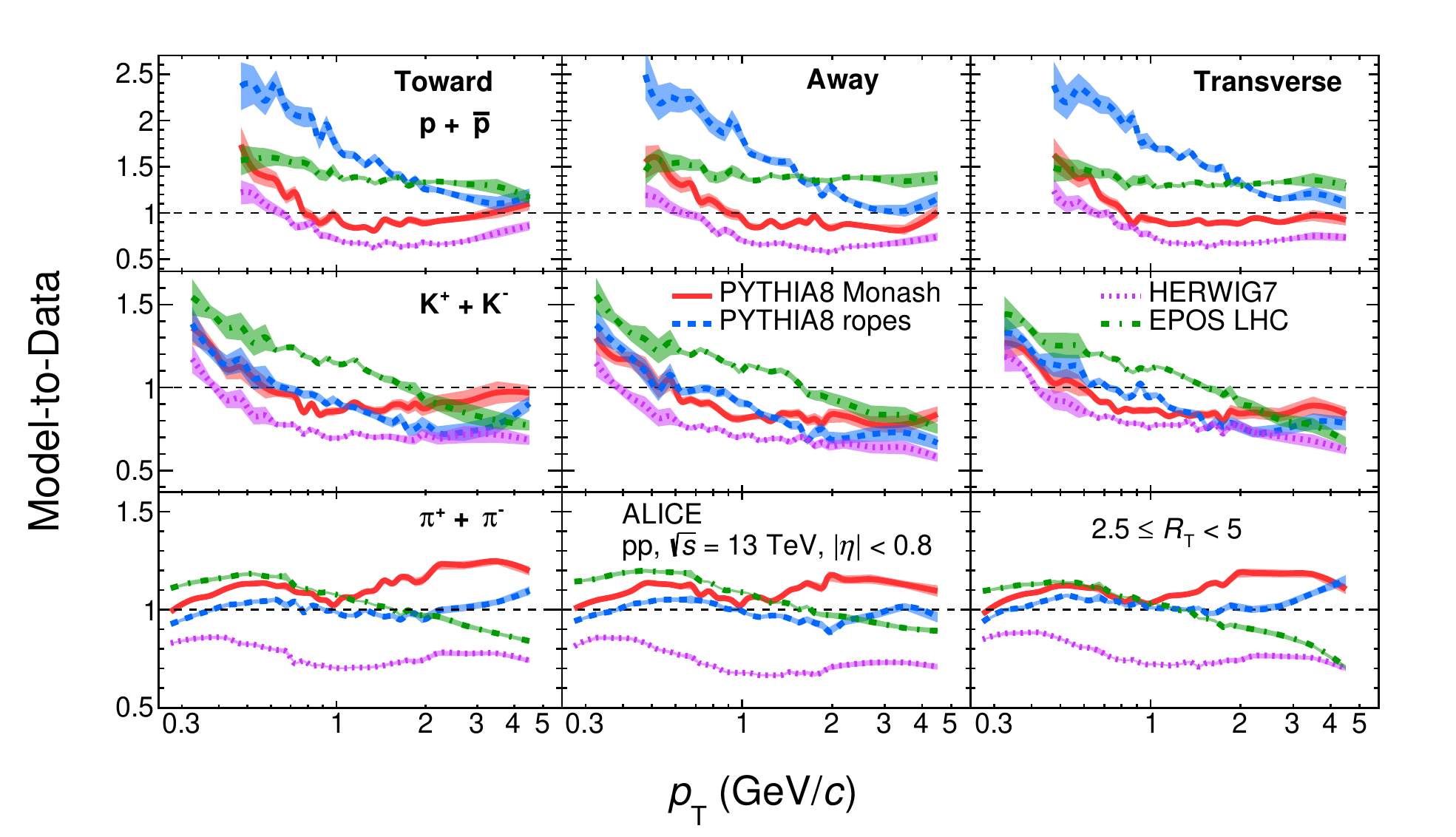}
    \hspace{0cm}
    \caption{Model-to-data ratios of the transverse momentum spectra. The results are shown for two \RT intervals: $0\leq \RT < 0.5$ (top figure) and $2.5\leq \RT < 5$ (bottom figure). The ratios in the toward, away and transverse regions are shown on the left, middle and right column, respectively. The error bands represent the combination of the statistical and systematic uncertainties on the model-to-data ratios.}
    \label{fig:Data_to_Model_spectra}
\end{figure}

Figure~\ref{fig:Data_to_Model_spectra} shows model-to-data ratios for the \pt spectra. The ratios are shown for two types of events: low UE activity $(0\leq \RT < 0.5)$ and high UE activity $(2.5\leq \RT < 5)$. It is observed that the models can describe the pion and kaon spectra for $\pt > 2~\gevc$ in the toward and away regions qualitatively for events with low UE activity. This is expected since for small \RT values, one mainly observes the jet fragmentation products, and the models are tuned to $\mathrm{e}^{+}\mathrm{e}^{-}$ data, which are jet-like. For this same \RT interval, the models predict different yields in the transverse region. However, for $\pt \gtrsim 1~\mathrm{GeV}/c$ all of the models underestimate the data. Moreover, increasing the UE activity makes the agreement between data and models worse. 

\clearpage

\begin{figure}[!ht]
    \centering
    \hspace{0cm}
    \includegraphics[width=1\textwidth]{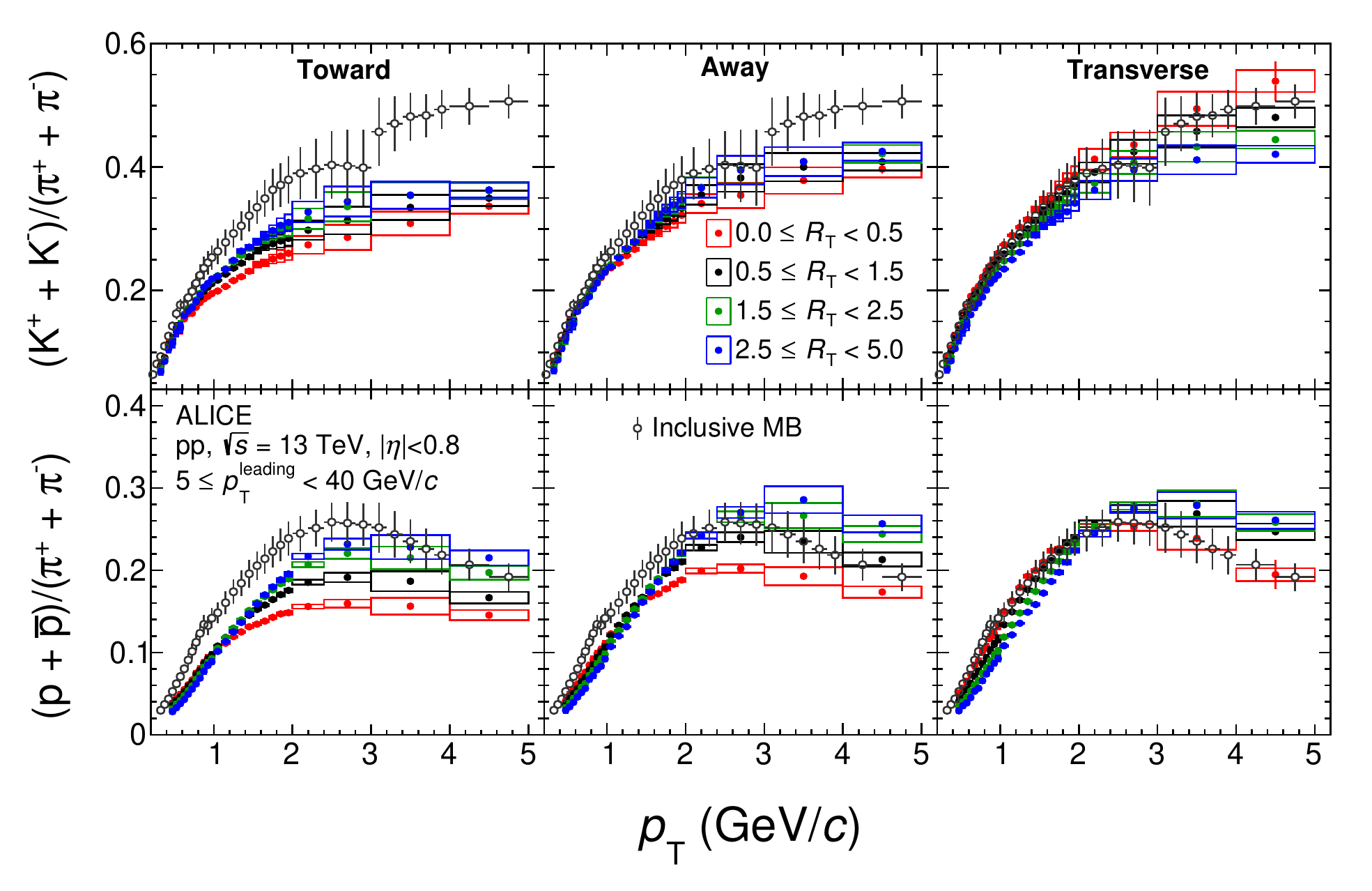}
    \caption{\pt-differential particle ratios as a function of \RT. The top (bottom) row shows the \ktopi (\ptopi) ratio. The results in the toward, away, and transverse regions are shown from left to right. Statistical and systematic uncertainties are represented with error bars and boxes, respectively. The inclusive minimum-bias particle ratios in \pp collisions at the same centre-of-mass energy~\cite{pikp_vs_mult_13TeV} are overlaid.}
    \label{fig:pT_Differential_Ratios_data}
\end{figure}

Figure~\ref{fig:pT_Differential_Ratios_data} shows the \pt-differential kaon-to-pion $(\ktopi)$ and proton-to-pion $(\ptopi)$ ratios for the four different \RT intervals in the three topological regions. The \RT-dependent ratios are contrasted with the inclusive ratios in minimum bias collisions at the same centre-of-mass energy~\cite{pikp_vs_mult_13TeV}. Minimum bias means integrated over \RT and the azimuthal angle, and without the leading particle requirement. The \ktopi ratios in the toward and away regions show similar features: they increase with increasing UE activity. However, this is true only for $1 \lesssim \pt < 2~\gevc$. Conversely, the \ktopi ratio in the transverse region decreases with increasing \RT. One also observes that the minimum-bias result is very similar to those measured in the transverse region. This suggests that the inclusive \ktopi ratio is dominated by bulk particle production. The \ptopi ratio in the toward and away regions measured in the lowest \RT intervals is always below the inclusive one. Similar observations have been made for the $\Lambda/\mathrm{K}^{0}_{\mathrm{S}}$ ratio in jets~\cite{Lambda_to_K0s_inJets}. As the UE increases, the toward and away regions become more UE dominated (jet dilution) and the \ptopi ratio also increases. However, this is true only for $\pt \gtrsim 1 ~\gevc$. The growth of the \ptopi ratio might be attributed to a gradual increase of the collective radial flow with \RT. Furthermore, the baryon-to-meson ratio for $\pt > 1~\mathrm{GeV}/c$ in these two regions tends to increase with increasing \RT and to approach the minimum bias ratio, which is similar to the one measured in the transverse region. The \ptopi ratio in the transverse region shows a mild dependence on \RT. It is observed that the result in the highest UE activity interval is below the one in the lowest UE activity interval for $\pt \lesssim 2~\gevc$, indicating a suppression of low-\pt protons possibly due to collective radial flow. Furthermore, the observed maximum in the highest \RT interval (centred at $\pt \approx 3.5~\gevc$) is shifted to the right with respect to the one of the lowest \RT interval (centred at $\pt \approx 2.5~\gevc$). This might be attributed to the jet hardening effect with increasing multiplicity as discussed in~\cite{Ortiz:2016kpz}.

\begin{figure}[!ht]
    \centering
    \hspace{0cm}
    \includegraphics[width=1\textwidth]{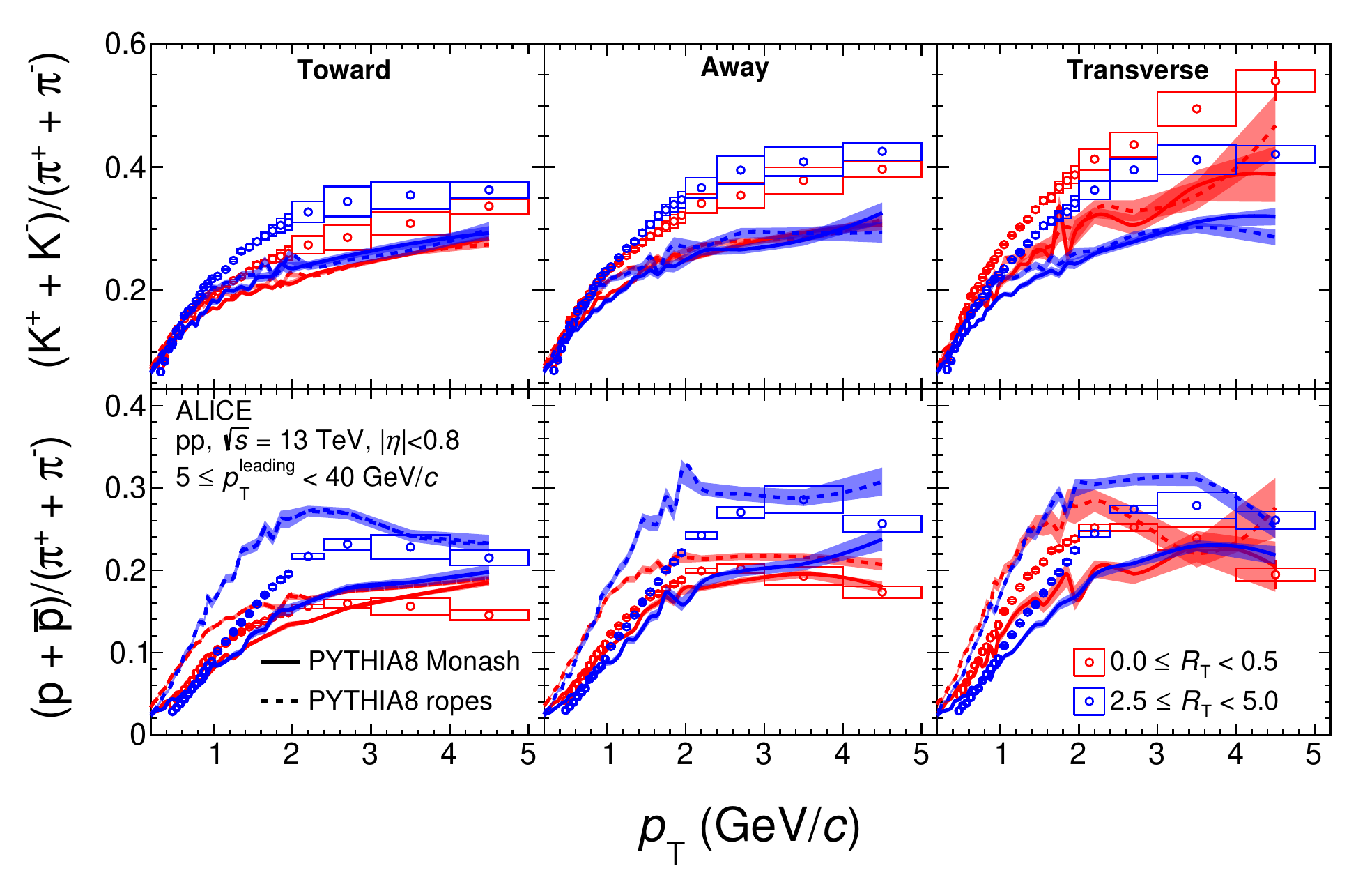}
    \hspace{0cm}
    \includegraphics[width=1\textwidth]{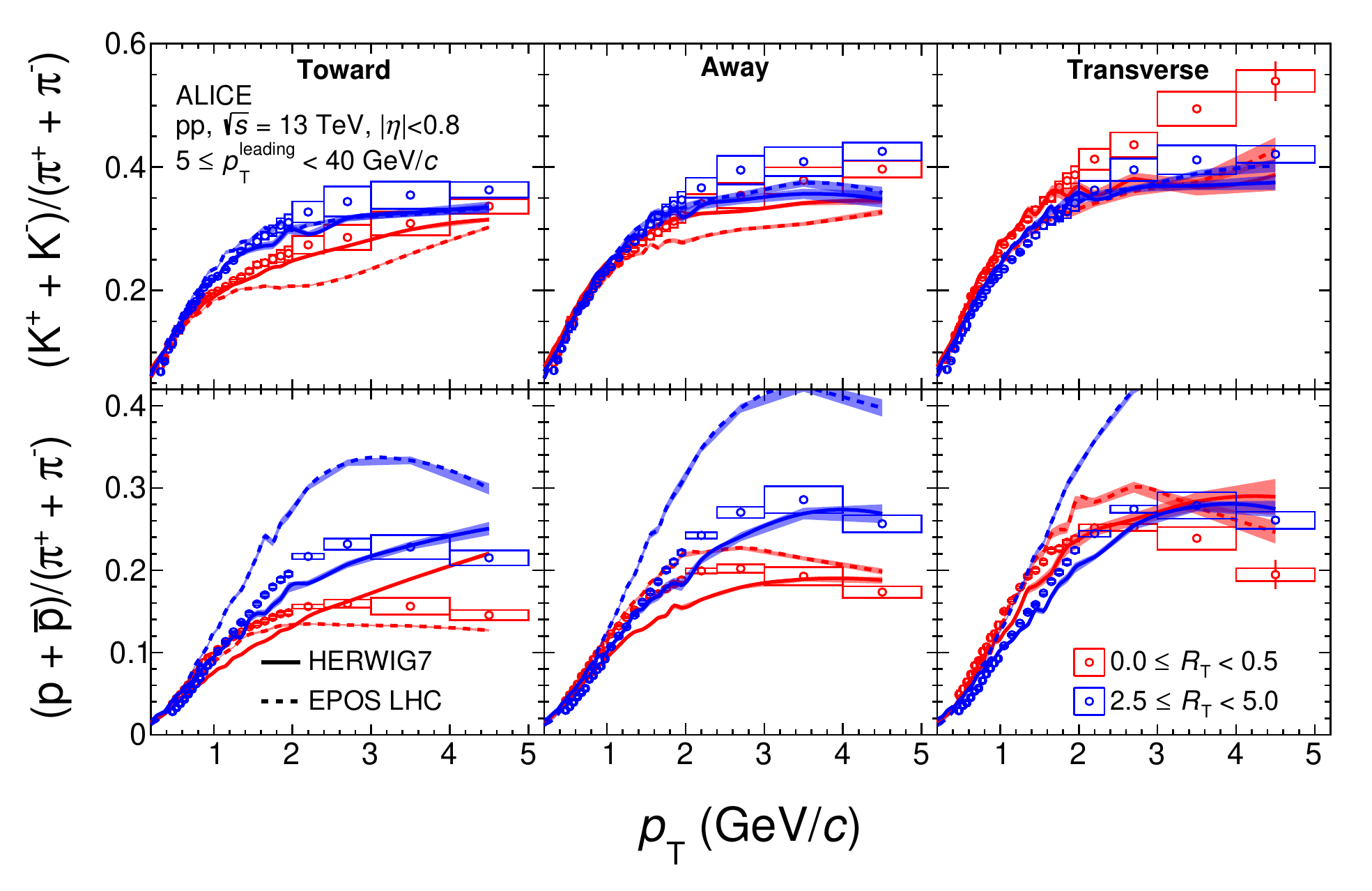}
    \caption{Kaon-to-pion and proton-to-pion ratios as a function of \pt for two \RT intervals: $0 \leq \RT < 0.5$ (red markers) and $2.5 \leq \RT < 5$ (blue markers). The particle ratios in the toward, away and, transverse regions are shown from left to right. The PYTHIA 8 Monash and PYTHIA 8 ropes (EPOS LHC and HERWIG7) predictions are shown in the top (bottom) figure. The shaded regions around the model line represent the statistical uncertainties.}
    \label{fig:pT_Differential_Ratios_data_wMCs}
\end{figure}

Figure~\ref{fig:pT_Differential_Ratios_data_wMCs} shows the \pt-differential \ktopi and \ptopi ratios along with model predictions in two \RT intervals: $0 \leq \RT < 0.5$ (low-UE activity) and  $2.5 < \RT < 5$ (high-UE activity). The \ktopi and \ptopi ratios in the toward and away regions in events at low \RT can be described qualitatively by PYTHIA8 Monash. However, this model predicts almost no evolution with \RT. On the other hand, the PYTHIA8 ropes hadronisation model, which allows for the formation of colour ropes, predicts \ptopi ratios that evolve with \RT, but overestimates the data, particularly for high-\RT events. EPOS LHC also describes both ratios qualitatively in the limit of low UE activity and predicts an evolution with \RT. It describes the \ktopi ratio but overestimates the \ptopi ratio in events with high \RT. This was clear from the \pt-integrated particle ratios: the transition from string fragmentation to statistical hadronisation needs improvement. Finally, HERWIG7 also predicts an evolution with \RT and can describe rather well the \ktopi ratio, while it misses the \pt trend of the \ptopi ratio. The fact that all models do a better job at describing both ratios at low than at high \RT is expected since they are tuned to $\mathrm{e}^{+}\mathrm{e}^{-}$ data. The model predictions in the away region are similar to those of the toward.

In the transverse region, PYTHIA8 Monash and PYTHIA8 ropes describe the splitting and ordering of the \ktopi ratio between the two \RT classes qualitatively but underestimate the data. They can also describe the \ptopi ratio qualitatively. Moreover, those models predict the lower \ptopi ratio for $\pt \lesssim 2~\gevc$ in events with high \RT compared to the low UE activity ratios. This effect, which can be attributed to the radial flow effects, is likely induced by the CR and ropes in PYTHIA8. EPOS LHC predicts the same \ktopi ratio for both \RT classes, while the \ptopi ratio at low \RT agrees with the data. Still, as previously mentioned, the transition from core-corona hadronisation is not well modeled. Finally, HERWIG7 gives a good qualitative description of the evolution of the \ptopi ratio with \RT in the transverse region.

The \pt-integrated yield $(\mathrm{d}N/\mathrm{d}y)$ and the average transverse momentum $(\meanpt)$ of pions, kaons, and protons are extracted from the \pt-differential spectra in the different \RT intervals and topological regions. Since the spectra are measured for $\pt > 0.3~\mathrm{GeV}/c~(\pion,\kaon)$ and $\pt > 0.45~(\pr)$~\gevc, they are first extrapolated to $\pt=0$. The extrapolation procedure is carried out by fitting the spectra with Lévy-Tsallis parameterisations~\cite{tsallis1988possible,Wilk:1999dr}. The parameterisation is only used in the \pt intervals with no data. For example, for the $0 \leq \RT < 0.5$ interval in the transverse region the fractions of extrapolated yields amount to $38\%$, $19\%$, and $22\%$ for \pion, \kaon, and \pr, respectively. To estimate the systematic uncertainty associated with the extrapolation procedure, several other parameterisations such as the Fermi-Dirac, Bose-Einstein, Blast-Wave, and $m_{\textrm{T}}$-exponential are used to estimate the extrapolated yield. The maximum difference between the nominal and extrapolated yields is associated as the systematic uncertainty of the extrapolation procedure. For example, the systematic uncertainties on the $\textrm{d}N/\textrm{d}y~(\meanpt)$ amount to $2\%(1.7\%)$, $2.7\%(2.3\%)$, and $2\%(1.5\%)$ for \pion, \kaon, and \pr, respectively, for the $0 \leq \RT < 0.5$ interval in the transverse region. 

Figure~\ref{fig:MeanpT_wMCs} shows the average transverse momentum as a function of \RT in the different topological regions. The \meanpt of \pion and \kaon in the toward region is the largest in the $0\leq \RT < 0.5$ (low UE activity) interval. This feature reflects the presence of the jet fragmenting mainly into low-mass hadrons (\pion and \kaon) with large transverse momentum. As the UE activity increases, the \meanpt of \pion and \kaon slowly decreases and tends to flatten for $\RT > 1.5$ due to the jet dilution effect: the toward and away regions become dominated by the UE. Conversely, the \meanpt of protons increases with \RT, which can be attributed to the additional radial flow effect. Moreover, the \meanpt of all the species at high-\RT tend to approach the values measured at high \RT in the transverse region. All models can describe the \meanpt qualitatively in the toward region, but EPOS LHC is the only one that predicts an increasing trend of the proton \meanpt. Particle production in the away region is similar to that in the toward. It is primarily dominated by the away-side jet. It is observed that the \meanpt of all the species increases with \RT. Furthermore, the \meanpt tends to approach the values of the transverse region at large \RT where the UE dominates. PYTHIA8 Monash and ropes give a fair qualitative description of the evolution of \meanpt of pions with \RT in the away region while the \meanpt of protons in the same region is only described by EPOS LHC. 

The \meanpt in the transverse region increases with \RT for all the species; however, the rate of increase exhibits a mass ordering, being more significant for heavier particles. Similar observations have been made in multiplicity-dependent studies~\cite{Mult_dependence_hadrons_7TeV_pp,pikp_vs_mult_13TeV}. The rise of the \meanpt with increasing \RT is likely attributed to autocorrelation effects. Since \RT and the \pt spectra are measured in the same $\Delta \varphi$ region, the high multiplicity requirement in the transverse region increases the probability to have a jet in the same region. Finally, it is observed that all the models predict the increase of the \meanpt with \RT.

\begin{figure}[!tb]
    \centering
    \hspace{0cm}
    \includegraphics[width=1\textwidth]{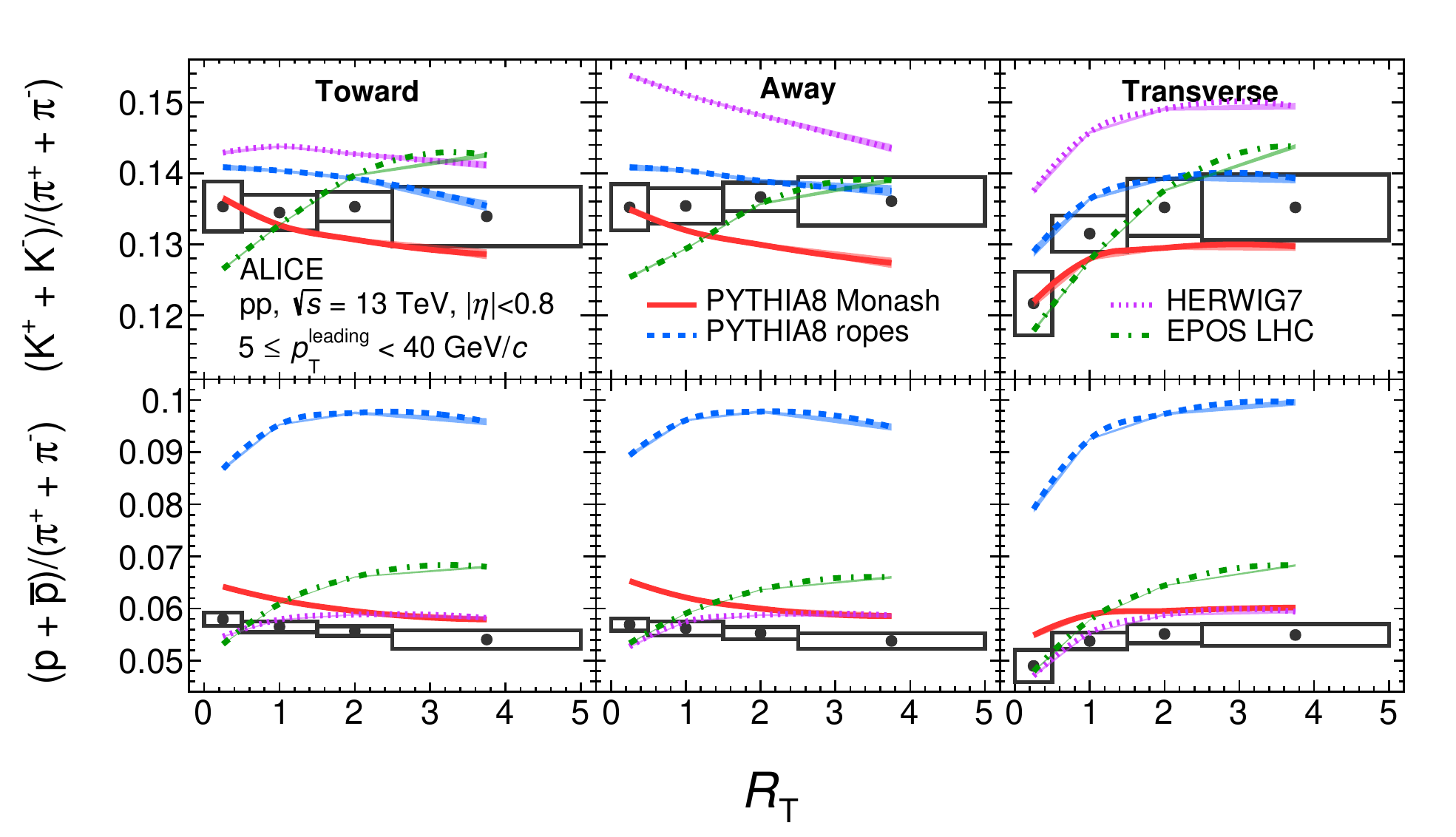}
    \caption{Transverse momentum-integrated particle ratios as a function of \RT. The particle ratios in the toward, away, and transverse regions are shown from left to right. The top (bottom) row plots the \ktopi (\ptopi). The statistical and systematic uncertainties are represented with error bars and boxes, respectively. The shaded bands around the model lines represent the statistical uncertainties.}
    \label{fig:IntegratedRatios_wMCs}
\end{figure}

Figure~\ref{fig:IntegratedRatios_wMCs} shows the \RT-dependence of the \pt-integrated particle ratios calculated from the extrapolated $\mathrm{d}N/\mathrm{d}y$. As the UE activity increases, the yield of kaons and protons relative to that of pions increases in the transverse region and in turn, the particle ratios grow until they saturate at $\RT \approx 1.5$. In contrast, in the toward and away regions the \ktopi ratio is constant as a function of \RT, while the \ptopi ratio decreases with increasing UE activity. Furthermore, both ratios in the toward and away regions approach the values of the transverse region at large \RT. All models predict the increasing trend of the particle ratios with \RT in the transverse region. PYTHIA8 Monash and HERWIG7 predict similar \ptopi ratios to the data. PYTHIA8 ropes overestimates the \ptopi by a large amount (almost a factor of 2) while EPOS LHC, although overpredicting, is closer to the data. One can notice that while EPOS LHC precisely describes the proton \meanpt as a function of \RT in all the topological regions, the \ptopi ratio is only described in the low-UE limit $(0\leq \RT <0.5)$, where string fragmentation dominates, indicating that the core overestimates the production of protons. PYTHIA8 Monash overpredicts the \ptopi ratio by about $10\,\%$ over the entire \RT range and underestimates the \ktopi ratio. In the toward and away regions none of the models reproduce the trend of both ratios: HERWIG7 and PYTHIA8 Monash predict a strongly decreasing trend for \ktopi that is not supported by the data. PYTHIA8 ropes and EPOS LHC predict an increasing trend for \ptopi that is in contradiction with the data. Finally, it is noted that while all the models capture most of the measured trends for \meanpt in Fig.~\ref{fig:MeanpT_wMCs}, none of the models describes the particle ratio trends of Fig.~\ref{fig:IntegratedRatios_wMCs} for both \ptopi and \ktopi.

%\clearpage

\section{Conclusions}
\label{sec:Conclusions}

The production of \pion, \kaon, and \pr was measured at mid-pseudorapidity in different topological regions as a function of the relative transverse activity classifier, \RT in \pp collisions at $\sqrt{s}=13~\mathrm{TeV}$ containing a high \pt $(5 \leq \pt^{\mathrm{leading}} < 40~\gevc)$ leading particle. \RT has been utilised to investigate differentially in different topological regions where the particle production is expected to be dominantly driven by pQCD-like processes (toward and away regions and low \RT) and regions where soft non-perturbative QCD processes dominate (transverse region or high \RT). In particular, since conventional UE (Underlying Event) studies average over the event activity, this analysis allows us to get further insight into collective effects and the interplay between hard and soft production in \pp collisions. Furthermore, the models can describe the new results in the toward and away regions when the UE is suppressed $(0 \leq \RT < 0.5)$, which was expected since they are tuned to reproduce jet-like $\mathrm{e}^{+}\mathrm{e}^{-}$ measurements. However, when the UE increases, all models fail to reproduce the data at both qualitative and quantitative level. This demonstrates that by measuring the production of identified particles as a function of \RT, one can reveal novel features of the UE. The new measurements presented here thus allow for substantial progress on the model side to nail down the properties of the UE.

%%%%%%%%%%%%%%%%%%%%%%%%%%%%%%%%
% end main text 
%%%%%%%%%%%%%%%%%%%%%%%%%%%%%%%%

%%%%% acknowledgements - handled by EB chairs 
\newenvironment{acknowledgement}{\relax}{\relax}
\begin{acknowledgement}
\section*{Acknowledgements}
% add specific acknowledgements here 
% ...but please don't remove the line below: funding agencies
% will be acknowledged with a custom tex file handled by EB chairs after Collab Round 2
% Version: 2022-11-04

The ALICE Collaboration would like to thank all its engineers and technicians for their invaluable contributions to the construction of the experiment and the CERN accelerator teams for the outstanding performance of the LHC complex.
The ALICE Collaboration gratefully acknowledges the resources and support provided by all Grid centres and the Worldwide LHC Computing Grid (WLCG) collaboration.
The ALICE Collaboration acknowledges the following funding agencies for their support in building and running the ALICE detector:
A. I. Alikhanyan National Science Laboratory (Yerevan Physics Institute) Foundation (ANSL), State Committee of Science and World Federation of Scientists (WFS), Armenia;
Austrian Academy of Sciences, Austrian Science Fund (FWF): [M 2467-N36] and Nationalstiftung f\"{u}r Forschung, Technologie und Entwicklung, Austria;
Ministry of Communications and High Technologies, National Nuclear Research Center, Azerbaijan;
Conselho Nacional de Desenvolvimento Cient\'{\i}fico e Tecnol\'{o}gico (CNPq), Financiadora de Estudos e Projetos (Finep), Funda\c{c}\~{a}o de Amparo \`{a} Pesquisa do Estado de S\~{a}o Paulo (FAPESP) and Universidade Federal do Rio Grande do Sul (UFRGS), Brazil;
Bulgarian Ministry of Education and Science, within the National Roadmap for Research Infrastructures 2020¿2027 (object CERN), Bulgaria;
Ministry of Education of China (MOEC) , Ministry of Science \& Technology of China (MSTC) and National Natural Science Foundation of China (NSFC), China;
Ministry of Science and Education and Croatian Science Foundation, Croatia;
Centro de Aplicaciones Tecnol\'{o}gicas y Desarrollo Nuclear (CEADEN), Cubaenerg\'{\i}a, Cuba;
Ministry of Education, Youth and Sports of the Czech Republic, Czech Republic;
The Danish Council for Independent Research | Natural Sciences, the VILLUM FONDEN and Danish National Research Foundation (DNRF), Denmark;
Helsinki Institute of Physics (HIP), Finland;
Commissariat \`{a} l'Energie Atomique (CEA) and Institut National de Physique Nucl\'{e}aire et de Physique des Particules (IN2P3) and Centre National de la Recherche Scientifique (CNRS), France;
Bundesministerium f\"{u}r Bildung und Forschung (BMBF) and GSI Helmholtzzentrum f\"{u}r Schwerionenforschung GmbH, Germany;
General Secretariat for Research and Technology, Ministry of Education, Research and Religions, Greece;
National Research, Development and Innovation Office, Hungary;
Department of Atomic Energy Government of India (DAE), Department of Science and Technology, Government of India (DST), University Grants Commission, Government of India (UGC) and Council of Scientific and Industrial Research (CSIR), India;
National Research and Innovation Agency - BRIN, Indonesia;
Istituto Nazionale di Fisica Nucleare (INFN), Italy;
Japanese Ministry of Education, Culture, Sports, Science and Technology (MEXT) and Japan Society for the Promotion of Science (JSPS) KAKENHI, Japan;
Consejo Nacional de Ciencia (CONACYT) y Tecnolog\'{i}a, through Fondo de Cooperaci\'{o}n Internacional en Ciencia y Tecnolog\'{i}a (FONCICYT) and Direcci\'{o}n General de Asuntos del Personal Academico (DGAPA), Mexico;
Nederlandse Organisatie voor Wetenschappelijk Onderzoek (NWO), Netherlands;
The Research Council of Norway, Norway;
Commission on Science and Technology for Sustainable Development in the South (COMSATS), Pakistan;
Pontificia Universidad Cat\'{o}lica del Per\'{u}, Peru;
Ministry of Education and Science, National Science Centre and WUT ID-UB, Poland;
Korea Institute of Science and Technology Information and National Research Foundation of Korea (NRF), Republic of Korea;
Ministry of Education and Scientific Research, Institute of Atomic Physics, Ministry of Research and Innovation and Institute of Atomic Physics and University Politehnica of Bucharest, Romania;
Ministry of Education, Science, Research and Sport of the Slovak Republic, Slovakia;
National Research Foundation of South Africa, South Africa;
Swedish Research Council (VR) and Knut \& Alice Wallenberg Foundation (KAW), Sweden;
European Organization for Nuclear Research, Switzerland;
Suranaree University of Technology (SUT), National Science and Technology Development Agency (NSTDA), Thailand Science Research and Innovation (TSRI) and National Science, Research and Innovation Fund (NSRF), Thailand;
Turkish Energy, Nuclear and Mineral Research Agency (TENMAK), Turkey;
National Academy of  Sciences of Ukraine, Ukraine;
Science and Technology Facilities Council (STFC), United Kingdom;
National Science Foundation of the United States of America (NSF) and United States Department of Energy, Office of Nuclear Physics (DOE NP), United States of America.
In addition, individual groups or members have received support from:
Marie Sk\l{}odowska Curie, European Research Council, Strong 2020 - Horizon 2020 (grant nos. 950692, 824093, 896850), European Union;
Academy of Finland (Center of Excellence in Quark Matter) (grant nos. 346327, 346328), Finland;
Programa de Apoyos para la Superaci\'{o}n del Personal Acad\'{e}mico, UNAM, Mexico.

\end{acknowledgement}

%%%%%%%% Bibliography 
\bibliographystyle{utphys}   % Remember we use title in the biblio
\bibliography{bibliography}
%\input {bibliography.tex}  

%%%%%%%%%%%%%%%%%%%%%%%%%%%%%%%%
% Appendices: yours (if any) + authorlist
%%%%%%%%%%%%%%%%%%%%%%%%%%%%%%%%
\newpage
\appendix

%
%\input{} % put your appendices here (if any)
%

%%%%% Authorlist - please do not touch: handled by EB chairs 
\section{The ALICE Collaboration}
\label{app:collab}
% ALICE Collaboration author list for 2022-11-04
\begin{flushleft} 
\small

S.~Acharya\,\orcidlink{0000-0002-9213-5329}\,$^{\rm 125}$, 
D.~Adamov\'{a}\,\orcidlink{0000-0002-0504-7428}\,$^{\rm 86}$, 
A.~Adler$^{\rm 69}$, 
G.~Aglieri Rinella\,\orcidlink{0000-0002-9611-3696}\,$^{\rm 32}$, 
M.~Agnello\,\orcidlink{0000-0002-0760-5075}\,$^{\rm 29}$, 
N.~Agrawal\,\orcidlink{0000-0003-0348-9836}\,$^{\rm 50}$, 
Z.~Ahammed\,\orcidlink{0000-0001-5241-7412}\,$^{\rm 132}$, 
S.~Ahmad\,\orcidlink{0000-0003-0497-5705}\,$^{\rm 15}$, 
S.U.~Ahn\,\orcidlink{0000-0001-8847-489X}\,$^{\rm 70}$, 
I.~Ahuja\,\orcidlink{0000-0002-4417-1392}\,$^{\rm 37}$
% ...

\section*{Affiliation Notes}

$^{\rm I}$ Deceased\\
$^{\rm II}$ Also at: Max-Planck-Institut f\"{u}r Physik, Munich, Germany\\
$^{\rm III}$ Also at: Italian National Agency for New Technologies, Energy and Sustainable Economic Development (ENEA), Bologna, Italy\\
$^{\rm IV}$ Also at: Dipartimento DET del Politecnico di Torino, Turin, Italy\\
$^{\rm V}$ Also at: Department of Applied Physics, Aligarh Muslim University, Aligarh, India\\
$^{\rm VI}$ Also at: Institute of Theoretical Physics, University of Wroclaw, Poland\\
$^{\rm VII}$ Also at: An institution covered by a cooperation agreement with CERN\\

\section*{Collaboration Institutes}

$^{1}$ A.I. Alikhanyan National Science Laboratory (Yerevan Physics Institute) Foundation, Yerevan, Armenia\\
$^{2}$ AGH University of Science and Technology, Cracow, Poland\\
$^{3}$ Bogolyubov Institute for Theoretical Physics, National Academy of Sciences of Ukraine, Kiev, Ukraine\\
$^{4}$ Bose Institute, Department of Physics  and Centre for Astroparticle Physics and Space Science (CAPSS), Kolkata, India\\
$^{5}$ California Polytechnic State University, San Luis Obispo, California, United States\\
$^{6}$ Central China Normal University, Wuhan, China\\
$^{7}$ Centro de Aplicaciones Tecnol\'{o}gicas y Desarrollo Nuclear (CEADEN), Havana, Cuba\\
$^{8}$ Centro de Investigaci\'{o}n y de Estudios Avanzados (CINVESTAV), Mexico City and M\'{e}rida, Mexico\\
$^{9}$ Chicago State University, Chicago, Illinois, United States\\
$^{10}$ China Institute of Atomic Energy, Beijing, China\\
$^{11}$ Chungbuk National University, Cheongju, Republic of Korea\\
$^{12}$ Comenius University Bratislava, Faculty of Mathematics, Physics and Informatics, Bratislava, Slovak Republic\\
$^{13}$ COMSATS University Islamabad, Islamabad, Pakistan\\
$^{14}$ Creighton University, Omaha, Nebraska, United States\\
$^{15}$ Department of Physics, Aligarh Muslim University, Aligarh, India\\
$^{16}$ Department of Physics, Pusan National University, Pusan, Republic of Korea\\
$^{17}$ Department of Physics, Sejong University, Seoul, Republic of Korea\\
$^{18}$ Department of Physics, University of California, Berkeley, California, United States\\
$^{19}$ Department of Physics, University of Oslo, Oslo, Norway\\
$^{20}$ Department of Physics and Technology, University of Bergen, Bergen, Norway\\
$^{21}$ Dipartimento di Fisica, Universit\`{a} di Pavia, Pavia, Italy\\
$^{22}$ Dipartimento di Fisica dell'Universit\`{a} and Sezione INFN, Cagliari, Italy\\
$^{23}$ Dipartimento di Fisica dell'Universit\`{a} and Sezione INFN, Trieste, Italy\\
$^{24}$ Dipartimento di Fisica dell'Universit\`{a} and Sezione INFN, Turin, Italy\\
$^{25}$ Dipartimento di Fisica e Astronomia dell'Universit\`{a} and Sezione INFN, Bologna, Italy\\
$^{26}$ Dipartimento di Fisica e Astronomia dell'Universit\`{a} and Sezione INFN, Catania, Italy\\
$^{27}$ Dipartimento di Fisica e Astronomia dell'Universit\`{a} and Sezione INFN, Padova, Italy\\
$^{28}$ Dipartimento di Fisica `E.R.~Caianiello' dell'Universit\`{a} and Gruppo Collegato INFN, Salerno, Italy\\
$^{29}$ Dipartimento DISAT del Politecnico and Sezione INFN, Turin, Italy\\
$^{30}$ Dipartimento di Scienze MIFT, Universit\`{a} di Messina, Messina, Italy\\
$^{31}$ Dipartimento Interateneo di Fisica `M.~Merlin' and Sezione INFN, Bari, Italy\\
$^{32}$ European Organization for Nuclear Research (CERN), Geneva, Switzerland\\
$^{33}$ Faculty of Electrical Engineering, Mechanical Engineering and Naval Architecture, University of Split, Split, Croatia\\
$^{34}$ Faculty of Engineering and Science, Western Norway University of Applied Sciences, Bergen, Norway\\
$^{35}$ Faculty of Nuclear Sciences and Physical Engineering, Czech Technical University in Prague, Prague, Czech Republic\\
$^{36}$ Faculty of Physics, Sofia University, Sofia, Bulgaria\\
$^{37}$ Faculty of Science, P.J.~\v{S}af\'{a}rik University, Ko\v{s}ice, Slovak Republic\\
$^{38}$ Frankfurt Institute for Advanced Studies, Johann Wolfgang Goethe-Universit\"{a}t Frankfurt, Frankfurt, Germany\\
$^{39}$ Fudan University, Shanghai, China\\
$^{40}$ Gangneung-Wonju National University, Gangneung, Republic of Korea\\
$^{41}$ Gauhati University, Department of Physics, Guwahati, India\\
$^{42}$ Helmholtz-Institut f\"{u}r Strahlen- und Kernphysik, Rheinische Friedrich-Wilhelms-Universit\"{a}t Bonn, Bonn, Germany\\
$^{43}$ Helsinki Institute of Physics (HIP), Helsinki, Finland\\
$^{44}$ High Energy Physics Group,  Universidad Aut\'{o}noma de Puebla, Puebla, Mexico\\
$^{45}$ Horia Hulubei National Institute of Physics and Nuclear Engineering, Bucharest, Romania\\
$^{46}$ Indian Institute of Technology Bombay (IIT), Mumbai, India\\
$^{47}$ Indian Institute of Technology Indore, Indore, India\\
$^{48}$ INFN, Laboratori Nazionali di Frascati, Frascati, Italy\\
$^{49}$ INFN, Sezione di Bari, Bari, Italy\\
$^{50}$ INFN, Sezione di Bologna, Bologna, Italy\\
$^{51}$ INFN, Sezione di Cagliari, Cagliari, Italy\\
$^{52}$ INFN, Sezione di Catania, Catania, Italy\\
$^{53}$ INFN, Sezione di Padova, Padova, Italy\\
$^{54}$ INFN, Sezione di Pavia, Pavia, Italy\\
$^{55}$ INFN, Sezione di Torino, Turin, Italy\\
$^{56}$ INFN, Sezione di Trieste, Trieste, Italy\\
$^{57}$ Inha University, Incheon, Republic of Korea\\
$^{58}$ Institute for Gravitational and Subatomic Physics (GRASP), Utrecht University/Nikhef, Utrecht, Netherlands\\
$^{59}$ Institute of Experimental Physics, Slovak Academy of Sciences, Ko\v{s}ice, Slovak Republic\\
$^{60}$ Institute of Physics, Homi Bhabha National Institute, Bhubaneswar, India\\
$^{61}$ Institute of Physics of the Czech Academy of Sciences, Prague, Czech Republic\\
$^{62}$ Institute of Space Science (ISS), Bucharest, Romania\\
$^{63}$ Institut f\"{u}r Kernphysik, Johann Wolfgang Goethe-Universit\"{a}t Frankfurt, Frankfurt, Germany\\
$^{64}$ Instituto de Ciencias Nucleares, Universidad Nacional Aut\'{o}noma de M\'{e}xico, Mexico City, Mexico\\
$^{65}$ Instituto de F\'{i}sica, Universidade Federal do Rio Grande do Sul (UFRGS), Porto Alegre, Brazil\\
$^{66}$ Instituto de F\'{\i}sica, Universidad Nacional Aut\'{o}noma de M\'{e}xico, Mexico City, Mexico\\
$^{67}$ iThemba LABS, National Research Foundation, Somerset West, South Africa\\
$^{68}$ Jeonbuk National University, Jeonju, Republic of Korea\\
$^{69}$ Johann-Wolfgang-Goethe Universit\"{a}t Frankfurt Institut f\"{u}r Informatik, Fachbereich Informatik und Mathematik, Frankfurt, Germany\\
$^{70}$ Korea Institute of Science and Technology Information, Daejeon, Republic of Korea\\
$^{71}$ KTO Karatay University, Konya, Turkey\\
$^{72}$ Laboratoire de Physique des 2 Infinis, Ir\`{e}ne Joliot-Curie, Orsay, France\\
$^{73}$ Laboratoire de Physique Subatomique et de Cosmologie, Universit\'{e} Grenoble-Alpes, CNRS-IN2P3, Grenoble, France\\
$^{74}$ Lawrence Berkeley National Laboratory, Berkeley, California, United States\\
$^{75}$ Lund University Department of Physics, Division of Particle Physics, Lund, Sweden\\
$^{76}$ Nagasaki Institute of Applied Science, Nagasaki, Japan\\
$^{77}$ Nara Women{'}s University (NWU), Nara, Japan\\
$^{78}$ National and Kapodistrian University of Athens, School of Science, Department of Physics , Athens, Greece\\
$^{79}$ National Centre for Nuclear Research, Warsaw, Poland\\
$^{80}$ National Institute of Science Education and Research, Homi Bhabha National Institute, Jatni, India\\
$^{81}$ National Nuclear Research Center, Baku, Azerbaijan\\
$^{82}$ National Research and Innovation Agency - BRIN, Jakarta, Indonesia\\
$^{83}$ Niels Bohr Institute, University of Copenhagen, Copenhagen, Denmark\\
$^{84}$ Nikhef, National institute for subatomic physics, Amsterdam, Netherlands\\
$^{85}$ Nuclear Physics Group, STFC Daresbury Laboratory, Daresbury, United Kingdom\\
$^{86}$ Nuclear Physics Institute of the Czech Academy of Sciences, Husinec-\v{R}e\v{z}, Czech Republic\\
$^{87}$ Oak Ridge National Laboratory, Oak Ridge, Tennessee, United States\\
$^{88}$ Ohio State University, Columbus, Ohio, United States\\
$^{89}$ Physics department, Faculty of science, University of Zagreb, Zagreb, Croatia\\
$^{90}$ Physics Department, Panjab University, Chandigarh, India\\
$^{91}$ Physics Department, University of Jammu, Jammu, India\\
$^{92}$ Physics Program and International Institute for Sustainability with Knotted Chiral Meta Matter (SKCM2), Hiroshima University, Hiroshima, Japan\\
$^{93}$ Physikalisches Institut, Eberhard-Karls-Universit\"{a}t T\"{u}bingen, T\"{u}bingen, Germany\\
$^{94}$ Physikalisches Institut, Ruprecht-Karls-Universit\"{a}t Heidelberg, Heidelberg, Germany\\
$^{95}$ Physik Department, Technische Universit\"{a}t M\"{u}nchen, Munich, Germany\\
$^{96}$ Politecnico di Bari and Sezione INFN, Bari, Italy\\
$^{97}$ Research Division and ExtreMe Matter Institute EMMI, GSI Helmholtzzentrum f\"ur Schwerionenforschung GmbH, Darmstadt, Germany\\
$^{98}$ Saga University, Saga, Japan\\
$^{99}$ Saha Institute of Nuclear Physics, Homi Bhabha National Institute, Kolkata, India\\
$^{100}$ School of Physics and Astronomy, University of Birmingham, Birmingham, United Kingdom\\
$^{101}$ Secci\'{o}n F\'{\i}sica, Departamento de Ciencias, Pontificia Universidad Cat\'{o}lica del Per\'{u}, Lima, Peru\\
$^{102}$ Stefan Meyer Institut f\"{u}r Subatomare Physik (SMI), Vienna, Austria\\
$^{103}$ SUBATECH, IMT Atlantique, Nantes Universit\'{e}, CNRS-IN2P3, Nantes, France\\
$^{104}$ Sungkyunkwan University, Suwon City, Republic of Korea\\
$^{105}$ Suranaree University of Technology, Nakhon Ratchasima, Thailand\\
$^{106}$ Technical University of Ko\v{s}ice, Ko\v{s}ice, Slovak Republic\\
$^{107}$ The Henryk Niewodniczanski Institute of Nuclear Physics, Polish Academy of Sciences, Cracow, Poland\\
$^{108}$ The University of Texas at Austin, Austin, Texas, United States\\
$^{109}$ Universidad Aut\'{o}noma de Sinaloa, Culiac\'{a}n, Mexico\\
$^{110}$ Universidade de S\~{a}o Paulo (USP), S\~{a}o Paulo, Brazil\\
$^{111}$ Universidade Estadual de Campinas (UNICAMP), Campinas, Brazil\\
$^{112}$ Universidade Federal do ABC, Santo Andre, Brazil\\
$^{113}$ University of Cape Town, Cape Town, South Africa\\
$^{114}$ University of Houston, Houston, Texas, United States\\
$^{115}$ University of Jyv\"{a}skyl\"{a}, Jyv\"{a}skyl\"{a}, Finland\\
$^{116}$ University of Kansas, Lawrence, Kansas, United States\\
$^{117}$ University of Liverpool, Liverpool, United Kingdom\\
$^{118}$ University of Science and Technology of China, Hefei, China\\
$^{119}$ University of South-Eastern Norway, Kongsberg, Norway\\
$^{120}$ University of Tennessee, Knoxville, Tennessee, United States\\
$^{121}$ University of the Witwatersrand, Johannesburg, South Africa\\
$^{122}$ University of Tokyo, Tokyo, Japan\\
$^{123}$ University of Tsukuba, Tsukuba, Japan\\
$^{124}$ University Politehnica of Bucharest, Bucharest, Romania\\
$^{125}$ Universit\'{e} Clermont Auvergne, CNRS/IN2P3, LPC, Clermont-Ferrand, France\\
$^{126}$ Universit\'{e} de Lyon, CNRS/IN2P3, Institut de Physique des 2 Infinis de Lyon, Lyon, France\\
$^{127}$ Universit\'{e} de Strasbourg, CNRS, IPHC UMR 7178, F-67000 Strasbourg, France, Strasbourg, France\\
$^{128}$ Universit\'{e} Paris-Saclay Centre d'Etudes de Saclay (CEA), IRFU, D\'{e}partment de Physique Nucl\'{e}aire (DPhN), Saclay, France\\
$^{129}$ Universit\`{a} degli Studi di Foggia, Foggia, Italy\\
$^{130}$ Universit\`{a} del Piemonte Orientale, Vercelli, Italy\\
$^{131}$ Universit\`{a} di Brescia, Brescia, Italy\\
$^{132}$ Variable Energy Cyclotron Centre, Homi Bhabha National Institute, Kolkata, India\\
$^{133}$ Warsaw University of Technology, Warsaw, Poland\\
$^{134}$ Wayne State University, Detroit, Michigan, United States\\
$^{135}$ Westf\"{a}lische Wilhelms-Universit\"{a}t M\"{u}nster, Institut f\"{u}r Kernphysik, M\"{u}nster, Germany\\
$^{136}$ Wigner Research Centre for Physics, Budapest, Hungary\\
$^{137}$ Yale University, New Haven, Connecticut, United States\\
$^{138}$ Yonsei University, Seoul, Republic of Korea\\
$^{139}$  Zentrum  f\"{u}r Technologie und Transfer (ZTT), Worms, Germany\\
$^{140}$ Affiliated with an institute covered by a cooperation agreement with CERN\\
$^{141}$ Affiliated with an international laboratory covered by a cooperation agreement with CERN.\\

\end{flushleft} 

\end{document}